\documentclass[journal]{IEEEtran}
\usepackage{cite,amssymb}
\usepackage[pdftex]{graphicx}
\usepackage{algorithm}
\usepackage[cmex10]{amsmath}
\usepackage{algorithm}
\usepackage{algorithm}
\usepackage{algpseudocode}
\usepackage{array,balance}

\ifCLASSOPTIONcompsoc
\usepackage[caption=false,font=normalsize,labelfont=sf,textfont=sf]{subfig}
\else
\usepackage[caption=false,font=footnotesize]{subfig}
\fi
\usepackage{url}
\usepackage{amsfonts}
\usepackage{exscale}
\usepackage{relsize}
\usepackage{multicol,color}

\newcommand{\mH}{\mathop{\rm H}}
\newcommand{\mT}{\mathop{\rm T}}
\begin{document}
	\title{OPARC: Optimal and Precise Array Response Control Algorithm -- Part I: Fundamentals}
	\author{Xuejing Zhang,~\IEEEmembership{Student Member,~IEEE,}
		Zishu He,~\IEEEmembership{Member,~IEEE,}
		Xiang-Gen Xia,~\IEEEmembership{Fellow,~IEEE,}\\
		Bin Liao,~\IEEEmembership{Senior Member,~IEEE,}
		Xuepan Zhang, and
		Yue Yang,~\IEEEmembership{Student Member,~IEEE}
		\thanks{X. Zhang, Z. He and Y. Yang are with the University of Electronic Science and Technology of China, Chengdu 611731, China (e-mail: xjzhang7@163.com; zshe@uestc.edu.cn; yueyang@std.uestc.edu.cn).}
		\thanks{X. Zhang and X.-G. Xia are
			with the Department of Electrical and Computer Engineering,
			University of Delaware, Newark, DE 19716, USA (e-mail: xjzhang@udel.edu; xxia@ee.udel.edu).}
		\thanks{B. Liao is with College of Information Engineering, Shenzhen University, Shenzhen 518060, China (e-mail: binliao@szu.edu.cn).}
		\thanks{X. P. Zhang is with Qian Xuesen Lab of Space Technology, Beijing 100094, China (e-mail: zhangxuepan@qxslab.cn).}}

\markboth{}
{Shell \MakeLowercase{\textit{et al.}}: Bare Demo of IEEEtran.cls for Journals}
\maketitle

\begin{abstract}
In this paper, the problem of how to optimally and precisely control array response levels is addressed.
By using the concept of the optimal weight vector from the adaptive array theory and adding
virtual interferences one by one, the change rule of the optimal weight vector
is found and a new formulation of the weight vector update is thus devised.
Then, the issue of how to precisely control the response level of one single direction
is investigated. More specifically, we assign a virtual interference to a direction such that
the response level
can be precisely controlled.
Moreover, the parameters, such as, the interference-to-noise ratio (INR), can be figured out according to the desired level.
Additionally, the parameter optimization is carried out to obtain the maximal array gain.
The resulting scheme is called optimal and precise array response control (OPARC)
in this paper.
To understand it better, its properties are given, and its comparison with the existing accurate array response control ($ {\textrm A}^2\textrm{RC} $) algorithm is provided.
Finally, simulation results are presented to verify the effectiveness and superiority of the
proposed OPARC.
\end{abstract}

\IEEEpeerreviewmaketitle

\section{Introduction}
\IEEEPARstart{A}{rray} antenna has been extensively applied in many fields, such as,
radar, navigation and wireless communications \cite{book}. It is known that the array pattern design is of significant importance to enhance system performance.
For instance, in radar systems, it is desirable to mitigate returns from interfering signals,
by designing a scheme which results in nulls at directions of interferences.
In some communication systems, it is critical to shape multiple-beam patterns for multi-user reception. Additionally, synthesizing a pattern with broad mainlobe is beneficial to extend monitoring areas in satellite remote sensing.
Generally speaking, array pattern can be designed either adaptively or non-adaptively.
Determining the complex weights for array elements so as to achieve a desired beampattern is known as array pattern synthesis. 
With regard to this problem,
it is expected to find weights that satisfy a set of specifications
on a given beampattern, in a data-independent or nonadaptive manner.

Over the past several decades, a great number of pattern synthesis approaches have been proposed, see, e.g., \cite{snrf06,snrf07,snrf08,snrf09,snrf11,snrf12,ref24,ref25,ref27,ref070, con3,con4,con5}.
Particularly, in \cite{con5}, an iterative sampling method is utilized to make
the sidelobe peaks conform to a specified shape within a given tolerance. For sidelobe control in cylindrical arrays, an artificially created noise source environment can be utilized \cite{con4}.
For a circular
ring array, a symmetrical pattern with low sidelobes is achieved in \cite{con3} by adopting a field-synthesis technique.
Note that although the solutions in \cite{con3,con4,con5} are able to control sidelobes
for arrays with some particular configurations, they cannot be straightforwardly extended to general geometries.
On the contrary, the accurate array response control ($ {\textrm A}^2\textrm{RC} $) approach \cite{snrf41}
provides a simple and effective manner to accurately control array response level of arbitrary arrays. Since the $ {\textrm A}^2\textrm{RC} $ approach deals with the response control of a single direction, more recently, a multi-point accurate
array response control (${\textrm{MA}}^2{\textrm{RC}}$)
method has been developed in \cite{2017-2} to flexibly adjust array responses of multiple points.
However, a satisfactory performance cannot be always guaranteed in either \cite{snrf41} or \cite{2017-2}
due to the empirical solution adopted in this kind of approaches.

These shortcomings of the existing approaches motivate us to have an innovative method to precisely, flexibly and optimally control the response level.
To do so, we first investigate how the optimal weight vector in the adaptive array theory changes along with the increase
of the number of interferences.
Then, a new scheme for weight vector update is developed
and further exploited to realize the precise array response control.
Furthermore, a parameter optimization mechanism is proposed
by maximizing the array gain \cite{refbookada1}.
It is shown that, the proposed optimal and precise array response control (OPARC) approach
is capable of precisely and flexibly controlling the response level of an arbitrary array.
Furthermore, its optimality (in the sense of array gain) can be well guaranteed.
In this paper, the proposed OPARC scheme is developed by exploiting the interference-to-noise ratio (INR).

It should be mentioned that this paper focuses on the main concepts and fundamentals of the OPARC scheme, while the extensions and applications
(such as pattern synthesis and quiescent pattern control \cite{con6}) will be carried out in the
companion paper \cite{p3}.
The rest of the paper is organized as follows.
Our proposed OPARC algorithm is presented in Section II.
Further insights into OPARC are presented in Section III
to provide more useful and interesting properties.
In Section IV,
comparisons between OPARC and the existing $ {\textrm A}^2\textrm{RC} $ are presented.
Representative simulations are conducted in Section V and conclusions are drawn in Section VI.

{\textit{Notations:}} We use bold upper-case and lower-case letters to represent
matrices and vectors, respectively.
In particular, we use $ {\bf I} $, $ {\bf 1} $ and $ {\bf 0} $ to denote the identity matrix,
the all-one vector and the all-zero vector, respectively.
$ j\triangleq\sqrt{-1} $.
$ (\cdot)^{\mT} $, $ (\cdot)^* $
and $ (\cdot)^{\mH} $ stand for the transpose, complex conjugate and Hermitian
transpose, respectively.
$ |\cdot| $ denotes the absolute value and $ \|\cdot\|_2 $ denotes the $ l_2 $ norm.
We use $ {\bf H}(i,l) $ to
stand for the element at the $ i $th row and $ l $th column of matrix $ {\bf H} $.
$ \Re(\cdot) $ and $ \Im(\cdot) $ denote the real
and imaginary parts, respectively.
$ {\rm det}(\cdot) $ is the determinant of a matrix.
The sign function is denoted by $ {\rm sign}(\cdot) $. $ \oslash $ represents the element-wise division operator.
We use $ {\rm diag}(\cdot) $ to
return a column vector composed of the diagonal elements of a matrix,
and use $ {\rm Diag}(\cdot) $ to stand for the diagonal matrix with the components of the input vector as the diagonal elements.
Finally, $ \mathbb{R} $ and $ \mathbb{C} $ denote the sets of all real and all complex numbers, respectively,
and $ \mathbb{S}^N_{++} $ denotes the set of $ N\times N $ positive definite matrices.

\section{$ \textrm{OPARC} $ Algorithm}
In order to present our proposed OPARC algorithm, we first briefly recall the adaptive array theory.
\subsection{Adaptive Array Theory}
Consider an array of $N$ elements and assume that the noise is white and the interferences are
independent with each other. To suppress the unwanted interferences and noise, the optimal adaptive beamformer weight vector
$\bf w$ steering to the direction $ \theta_0 $ can be obtained by maximizing the output signal-to-interference-plus-noise ratio (SINR) defined as
\begin{align}\label{defSINR}
{\rm SINR}=\dfrac{{\sigma}^2_s|{\bf w}^{\mH}{\bf a}(\theta_0)|^2}{{\bf w}^{\mH}{\bf R}_{n+i}{\bf w}}
\end{align}
where $ {\sigma}^2_s $ stands for the
signal power, $ {\bf R}_{n+i} $ denotes the $N\times N$ noise-plus-interference covariance matrix
and $ {\bf a}(\theta_0) $ represents the signal steering vector.
More exactly, for a given $ \theta $, we have
\begin{align}\label{atheta}
{\bf a}(\theta)=[ g_1(\theta)e^{-j\omega\tau_1(\theta)},\cdots,g_N(\theta)e^{-j\omega\tau_N(\theta)}]^{\mT}
\end{align}
where $ g_n(\theta) $ denotes the pattern of the $n$th element, $ \tau_n(\theta) $ is the time-delay between the $n$th element and 
the reference point, $ n=1,\cdots,N $, $ \omega $ denotes the operating frequency.

It is known that the optimal weight vector $ {\bf w}_{\rm opt} $, which maximizes the SINR, is given by \cite{refbookada1}
\begin{align}\label{eqn0001}
{\bf w}_{\rm opt}=\alpha{\bf R}^{-1}_{n+i}{\bf a}(\theta_0)
\end{align}
where $ \alpha $ is a normalization factor and does not affect the output SINR, and hence, will be omitted in the sequel.

Note that the above SINR can be expressed as $G\cdot{\sigma_s^2}/{\sigma_n^2} $, where
$G$ is defined as
\begin{align}\label{Gdef}
G=\dfrac{|{\bf w}^{\mH}{\bf a}(\theta_0)|^2}{{\bf w}^{\mH}{\bf T}_{n+i}{\bf w}}
\end{align}
with ${\bf T}_{n+i}\triangleq{{\bf R}_{n+i}}/{\sigma^2_n}$ standing for the normalized noise-plus-interference covariance matrix, i.e.,
\begin{align}\label{eqn1004}
{\bf T}_{n+i}=\frac{{\bf R}_{n+i}}{\sigma_n^2}={\bf I}+\sum_{\ell=1}^{k}{\beta}_\ell{\bf a}(\theta_\ell){\bf a}^{\mH}(\theta_\ell)
\end{align}
where $ {\beta}_\ell\triangleq{{\sigma}^2_\ell}/{{\sigma}^2_n} $ denotes the interference-to-noise ratio (INR), $ k $ is the number of interferences,
$ {\bf a}(\theta_\ell) $ is the steering vector of the $ \ell $th interference,
$ {\sigma}^2_n $ and $ {\sigma}^2_\ell $ stand for the noise and interference powers, respectively.

Note that $G$ represents the amplification factor of the input signal-to-noise ratio
(SNR) $ {\sigma_s^2}/{\sigma_n^2} $, and therefore, is termed as the array gain \cite{refbookada1}. As a result, the criterion of  array gain $G$ maximization is adopted to achieve the optimal weight vector.

\subsection{Update of the Optimal Weight Vector}
It can be seen from \eqref{eqn0001}--\eqref{eqn1004} that the optimal weight vector $ {\bf w}_{\rm opt} $ depends on $ {{\bf R}_{n+i}} $ or $ {\bf T}_{n+i} $, which is not available for the following data-independent array response control: for a given steering vector $ {\bf a}(\theta) $ in \eqref{atheta} and a beam axis $ \theta_0 $, design a weight vector $\bf w$ such that the normalized array response $L(\theta,\theta_0)\triangleq{|{\bf w}^{\mH}{\bf a}(\theta)|^2}\big/{|{\bf w}^{\mH}{\bf a}(\theta_0)|^2}$ meets some specific requirements. 
In	this paper, we are interested in the requirements of array response levels,
i.e., finding weight vectors such that the array responses at a given set of angles are equal to a set of predescribed values.
Our basic idea is to construct a virtual normalized noise-plus-interference covariance matrix (VCM), denoted as $ {\bf T}_k $, to 
achieve the given response control task. Note that since the VCM $ {\bf T}_k $ to be determined is not produced by real data, it may not have any physical meaning. Moreover, it can be neither positive definite nor Hermitian (its rationality will be discussed later). By making use of the VCM, the data-dependent adaptive array theory can be applied to the data-independent situation considered in this paper. This allows us to optimally update the weight vector $ {\bf w}_{k-1,{\rm opt}}={\bf T}^{-1}_{k-1}{\bf a}(\theta_0)$ to $ {\bf w}_{k,{\rm opt}}$ such that a desired response level $ \rho_k $ at $ \theta_k $ can be achieved by assigning an appropriate virtual interference. Thus, the problem we concern here is to figure out the characteristics, e.g., INR, of the virtual interference.

%
%

We use induction to describe the problem and the algorithm below.
Suppose that the response levels of the $ k-1 $ directions have been successively controlled by adding $ k-1 $ virtual interferences. Meanwhile, the corresponding VCM is denoted as
$ {\bf T}_{k-1} $.
For a given $ \theta_k $ and its desired level $ \rho_k $, we can assign the $ k $th
virtual interference coming from $ \theta_k $ by designing its INR (i.e.,
$ {\beta_{k}} $). To find out $ {\beta_{k}} $, from \eqref{eqn1004} we notice that the VCM can be updated as
\begin{align}\label{modi0018}
{\bf T}_k={\bf T}_{k-1}+{\beta}_k{\bf a}(\theta_k){\bf a}^{\mH}(\theta_k).
\end{align}
Using the Woodbury Lemma \cite{refmat}, we have
\begin{align}\label{modi0008}
{\bf T}^{-1}_{k}
={\bf T}^{-1}_{k-1}-
\dfrac{\beta_k{\bf T}^{-1}_{k-1}{\bf a}(\theta_k){\bf a}^{\mH}(\theta_k){\bf T}^{-1}_{k-1}}
{1+{\beta_k}{\bf a}^{\mH}(\theta_k){\bf T}^{-1}_{k-1}{\bf a}(\theta_k)}.
\end{align}
Accordingly, the optimal weight vector
is given by ${\bf w}_{k,{\rm opt}}={\bf T}^{-1}_{k}{\bf a}(\theta_0)$.
Recalling \eqref{eqn0001} and \eqref{modi0008}, we can express $ {\bf w}_{k,{\rm opt}} $ as
\begin{align}\label{modi0019}
{\bf w}_{k,{\rm opt}}={\bf w}_{k-1,{\rm opt}}+{\gamma}_k{\bf T}^{-1}_{k-1}{\bf a}(\theta_k)
\end{align}
where $ {\bf w}_{k-1,{\rm opt}}={\bf T}^{-1}_{k-1}{\bf a}(\theta_0) $ denotes
the previous optimal weight vector
and $ {\gamma}_k $ is given by
\begin{align}\label{modi0005}
{\gamma}_k=-
\dfrac{\beta_k{\bf a}^{\mH}(\theta_k){\bf T}^{-1}_{k-1}{\bf a}(\theta_0)}
{1+{\beta_k}{\bf a}^{\mH}(\theta_k){\bf T}^{-1}_{k-1}{\bf a}(\theta_k)}
\triangleq\Psi_{k}({\beta_k})
\end{align}
with $ \Psi_{k}(\cdot) $ denoting a mapping from $ {\beta_k} $
to $ {\gamma}_k $.

Note that the solution in \eqref{modi0019}--\eqref{modi0005} gives the optimal solution for maximizing the SINR, which may not meet the response level $\rho_k$ at $\theta_k$. In order to meet this response level requirement, we next consider the following questions first. Given the previous weight vector $ {\bf w}_{k-1,{\rm opt}}={\bf T}^{-1}_{k-1}{\bf a}(\theta_0) $,
does there exist $ \gamma_k $ (or equivalently $ \beta_k $) such that the response level at $ \theta_k $ is precisely
$\rho_k$? and what value it should be if it exists?
To do so, we reformulate the weight vector as
\begin{align}\label{modi0020}
{\bf w}_k={\bf w}_{k-1}+{\gamma}_k{\bf v}_{k}
\end{align}
where the subscript $ (\cdot)_{\rm opt} $ is omitted for notational simplicity and $ {\bf v}_k $ is defined as
\begin{align}\label{modi0021}
{\bf v}_k\triangleq{\bf T}^{-1}_{k-1}{\bf a}(\theta_k).
\end{align}

Mathematically, the problem of finding $ {\gamma}_k $ such that the array response level at $\theta_k$ is $\rho_k$ can be written as
\begin{align}\label{Ldef}
  L(\theta_k,\theta_0)={|{\bf w}^{\mH}_{k}{\bf a}(\theta_k)|^2}\big/{|{\bf w}^{\mH}_{k}{\bf a}(\theta_0)|^2}={\rho_k}
\end{align}
where the desired array response level satisfies  $ \rho_k \le 1$.
The combination of \eqref{modi0020} and \eqref{Ldef} yields
\begin{align}\label{modi0025}
{\bf z}^{\mH}_k {\bf H}_k {\bf z}_k=0
\end{align}
where ${\bf z}_k$ and $ {\bf H}_k $ are, respectively, defined as
\begin{equation}\label{eqn0013}
\begin{split}
{\bf z}_k&\triangleq[1 ~~\gamma_k]^{\mT}\\
	{\bf H}_k\!&\triangleq\![{\bf w}_{k\!-\!1}~{\bf v}_k]^{\mH} \!	\left(\! 	{\bf a}(\!\theta_k\!) {\bf a}^{\mH}(\!\theta_k\!)\!\!-\!\!\rho_k{\bf a}(\!\theta_0\!){\bf a}^{\mH}(\!\theta_0\!)\!\right)\! [	{\bf w}_{k\!-\!1}\!~{\bf v}_k].\\
\end{split}
\end{equation}
By expanding \eqref{modi0025} and \eqref{eqn0013}, we immediately have the following proposition.
\newtheorem{theorem}{Proposition}
\begin{theorem}
	Suppose that $ \gamma_k $ (i.e., the second entry of $ {\bf z}_k $) satisfies \eqref{modi0025},
	if $ {\bf H}_k(2,2)=0 $, it can be
	derived that the trajectory of
	$ \begin{bmatrix}	{\Re}(\gamma_k)& {\Im}(\gamma_k)\end{bmatrix}^{\mT}  $ is a line as
	\begin{align}
	{\Re}[{\bf H}_k(1,2)]{\Re}({\gamma}_k)-{\Im}[{\bf H}_k(1,2)]{\Im}({\gamma}_k)
	=-{{\bf H}_k(1,1)}/{2}.\nonumber
	\end{align}	
	If $ {\bf H}_k(2,2)\neq0 $, the trajectory of
	$ [	{\Re}(\gamma_k)~~ {\Im}(\gamma_k)]^{\mT}  $ is a
	circle, denoted by $ \mathbb{C}_{\gamma} $:
	\begin{align}
	\mathbb{C}_{\gamma}=\left\{
	[	{\Re}(\gamma_k)~~ {\Im}(\gamma_k)]^{\mT}\Big|
	\big\|[	{\Re}(\gamma_k)~~ {\Im}(\gamma_k)]^{\mT}-{\bf c}_{\gamma}\big\|_2=R_{\gamma}
	\right\}\nonumber
	\end{align}	
	 with the center
	\begin{align}\label{eqn0030}
	{\bf c}_{\gamma}=\dfrac{1}{{\bf H}_k(2,2)}
	\begin{bmatrix}-{\Re}\left[{\bf H}_k(1,2)\right]\\ {\Im}\left[{\bf H}_k(1,2)\right]\end{bmatrix}
	\end{align}
	and the radius
	\begin{align}\label{eqn0031}
	R_{\gamma}={\sqrt{-{\rm det}({\bf H}_k)}}\big/{|{\bf H}_k(2,2)|}.
	\end{align}
\end{theorem}

From this proposition, it is known that, given the previous weight vector $ {\bf w}_{k-1}={\bf T}^{-1}_{k-1}{\bf a}(\theta_0) $, there exist infinitely many solutions of $ \gamma_k $ to
achieve a response level of $ \rho_k $ at $ \theta_k $. This implies that the response level at a certain direction can be precisely adjusted by assigning a virtual interference with properly designed INR parameter $\gamma_k$.

It is clear that $ {\bf H}_k(2,2)=0 $ is equivalent to
\begin{align}\label{barrho}
{\rho}_k=\dfrac{\big|
	({\bf T}^{-1}_{k-1}{\bf a}(\theta_k))^{\mH}{\bf a}(\theta_k)
	\big|^2}
{\big|
	({\bf T}^{-1}_{k-1}{\bf a}(\theta_k))^{\mH}{\bf a}(\theta_0)
	\big|^2}.
\end{align}
In this case $ {\rho}_{k} $ is equal to the normalized
power response at $ \theta_k $ when the weight vector is $ {\bf T}^{-1}_{k-1}{\bf a}(\theta_k) $,
i.e., when the beampattern steers to the beam axis $ \theta_k $.
Typically, a beampattern reaches its maximum at the beam axis, i.e., we have $ {\rho}_{k}> 1 $ if $\theta_k \neq \theta_0$. This
would contradict with the fact that $\rho_k \le 1$. Hence, $ {\bf H}_k(2,2)=0 $ usually will not occur and in the sequel we only focus on the case of $ {\bf H}_k(2,2)\neq0 $,
and from Proposition 2, the trajectory of
$ [	{\Re}(\gamma_k)~ {\Im}(\gamma_k)]^{\mT}  $ is a circle, as illustrated in Fig. \ref{controlall}. Then, the remaining question is among all these valid solutions of $\gamma_k$ (or $\beta_k$), to meet the response level requirement, which one is to maximize the SINR or beam gain? We will study this question below.

\subsection{Selection of $ {\gamma}_k $ and Update of the Weight Vector}
The preceding problem can be formulated as the following constrained optimal and precise array response control (OPARC) problem:
\begin{subequations}\label{qu011}	
	\begin{align}
	\label{qu5210}\mathop {{\rm{maximize}}}\limits_{{\gamma _k}} &~~~G_k
	\triangleq{|{\bf w}^{\mH}_k{\bf a}(\theta_0)|^2}/{|{\bf w}^{\mH}_k{\bf T}_k{\bf w}_k|}\\
	\label{qu5211}{\rm subject~to}&~~~L(\theta_k,\theta_0)={\rho_k}\\
	\label{qu5212}&~~~{\bf w}_k={\bf w}_{k-1}+{\gamma}_k{\bf T}^{-1}_{k-1}{\bf a}(\theta_k).
	\end{align}
\end{subequations}
From \eqref{qu5210}, one can see that the desired weight vector is expected to provide the maximum array
gain with some additional constraints. Moreover, in the above devised OPARC scheme, apart from the response level constraint \eqref{qu5211}, we have also
imposed constraint \eqref{qu5212} to make the resultant $ {\bf w}_k $ be
a particular optimal weight vector in correspondence to assigning another
interference at $ \theta_k $ to the existing $k-1$ interferences at directions $\theta_1,\ldots,\theta_{k-1}$.

When $ {\gamma}_k $ satisfies \eqref{modi0005}, \eqref{qu5212} leads to $ {\bf w}_k={\bf T}^{-1}_{k}{\bf a}(\theta_0)$.
In order to solve problem \eqref{qu011}, we first substitute $ {\bf w}_k={\bf T}^{-1}_{k}{\bf a}(\theta_0)$ into the objective function and get $G_k=|{\bf a}^{\mH}(\theta_0){\bf T}^{-\mH}_{k}{\bf a}(\theta_0)|$. Then, recalling (7) and (9), we can rewrite $G_k$ as
\begin{align}\label{defG}
G_k&=\big|{\bf a}^{\mH}(\theta_0){\bf T}^{-1}_{k-1}{\bf a}(\theta_0)+
{\gamma}_k{\bf a}^{\mH}(\theta_0){\bf T}^{-1}_{k-1}{\bf a}(\theta_k)\big|\nonumber\\
&=|\widetilde{{\xi}}_c|\cdot
|{{\xi}_0}/{\widetilde{{\xi}}_c}+{\gamma}_k|
\end{align}
where ${\xi}_0$, ${\xi}_k$, ${\xi}_c$ and $\widetilde{{\xi}}_c$ are defined as
\begin{subequations}\label{eqn059}
	\begin{align}
	{\xi}_0&\triangleq{\bf a}^{\mH}(\theta_0){\bf T}^{-1}_{k-1}{\bf a}(\theta_0)\\
\label{20b}{\xi}_k&\triangleq{\bf a}^{\mH}(\theta_k){\bf T}^{-1}_{k-1}{\bf a}(\theta_k)\\
	{\xi}_c&\triangleq{\bf a}^{\mH}(\theta_k){\bf T}^{-1}_{k-1}{\bf a}(\theta_0)\\
	\widetilde{{\xi}}_c&\triangleq{\bf a}^{\mH}(\theta_0){\bf T}^{-1}_{k-1}{\bf a}(\theta_k).
	\end{align}
\end{subequations}
Then, from Proposition 1, problem \eqref{qu011} can be expressed as
\begin{subequations}\label{qu311}	
	\begin{align}
	\mathop {{\rm{maximize}}}\limits_{{\gamma _k}}&~~~|{{\xi}_0}/{\widetilde{{\xi}}_c}+{\gamma}_k|\\
\label{qu312}{\rm subject~to}&~~~
	\begin{bmatrix}	{\Re}(\gamma_k)& {\Im}(\gamma_k)\end{bmatrix}^{\mT}\in\mathbb{C}_{\gamma}.
	\end{align}
\end{subequations}
Although the problem \eqref{qu311} is non-convex, it will be shown
that it can be
analytically solved as follows.
\begin{theorem}
Denote the intersections of the circle and the line connecting the origin $ {\bf O}=[0,0]^{\mT} $ and
the center $ {\bf c}_{\gamma} $ in \eqref{eqn0030} as $ F_a\triangleq[{\Re}({\gamma}_{k,a})~{\Im}({\gamma}_{k,a})]^{\mT} $
and $ F_b\triangleq[{\Re}({\gamma}_{k,b})~{\Im}({\gamma}_{k,b})]^{\mT} $, respectively,
and assume that $ |{\gamma}_{k,a}|<|{\gamma}_{k,b}| $. If $ {\bf T}_{k-1} $ is Hermitian
(note that the Hermitian property of $ {\bf T}_{k-1} $
has not been guaranteed as we have mentioned earlier), then the optimal solution of \eqref{qu311}
satisfies
\begin{align}\label{eqn0361}
{\gamma}_{k,\star}=\left\{
\begin{array}{cc}
{\gamma}_{k,a},&{\textrm{if}}~\zeta>0\\
{\gamma}_{k,b},&{\textrm{otherwise}}
\end{array} \right.
\end{align}
where
\begin{align}\label{zetadef}
\zeta\triangleq{\rm sign}[{\bf c}_{\gamma}(1)]\cdot{\rm sign}[{\Re}(d)-{\bf c}_{\gamma}(1)]
\end{align}
and
\begin{align}\label{defd}
d\triangleq-{{\xi}_0}\big/{\xi}^{*}_c.
\end{align}
In addition, $ {\gamma}_{k,a} $ and $ {\gamma}_{k,b} $ in \eqref{eqn0361} are calculated as
\begin{align}
{\gamma}_{k,a}=-\dfrac{\left(\|{\bf c}_{\gamma}\|_2-R_{\gamma}\right)\chi{\xi}_c}
{\|{\bf c}_{\gamma}\|_2{\bf H}_k(2,2)},~
{\gamma}_{k,b}=-\dfrac{\left(\|{\bf c}_{\gamma}\|_2+R_{\gamma}\right)\chi{\xi}_c}
{\|{\bf c}_{\gamma}\|_2{\bf H}_k(2,2)}\nonumber
\end{align}
where $ \chi={\xi}_k-{\rho_k}{\xi}_0\in\mathbb{R}$, $ {\bf c}_{\gamma} $ and $ R_{\gamma} $ are defined in Proposition 1.
\end{theorem}
\begin{IEEEproof}
	See Appendix A.	
\end{IEEEproof}

\begin{figure}[!t]
	\centering
	\includegraphics[width=3.0in]{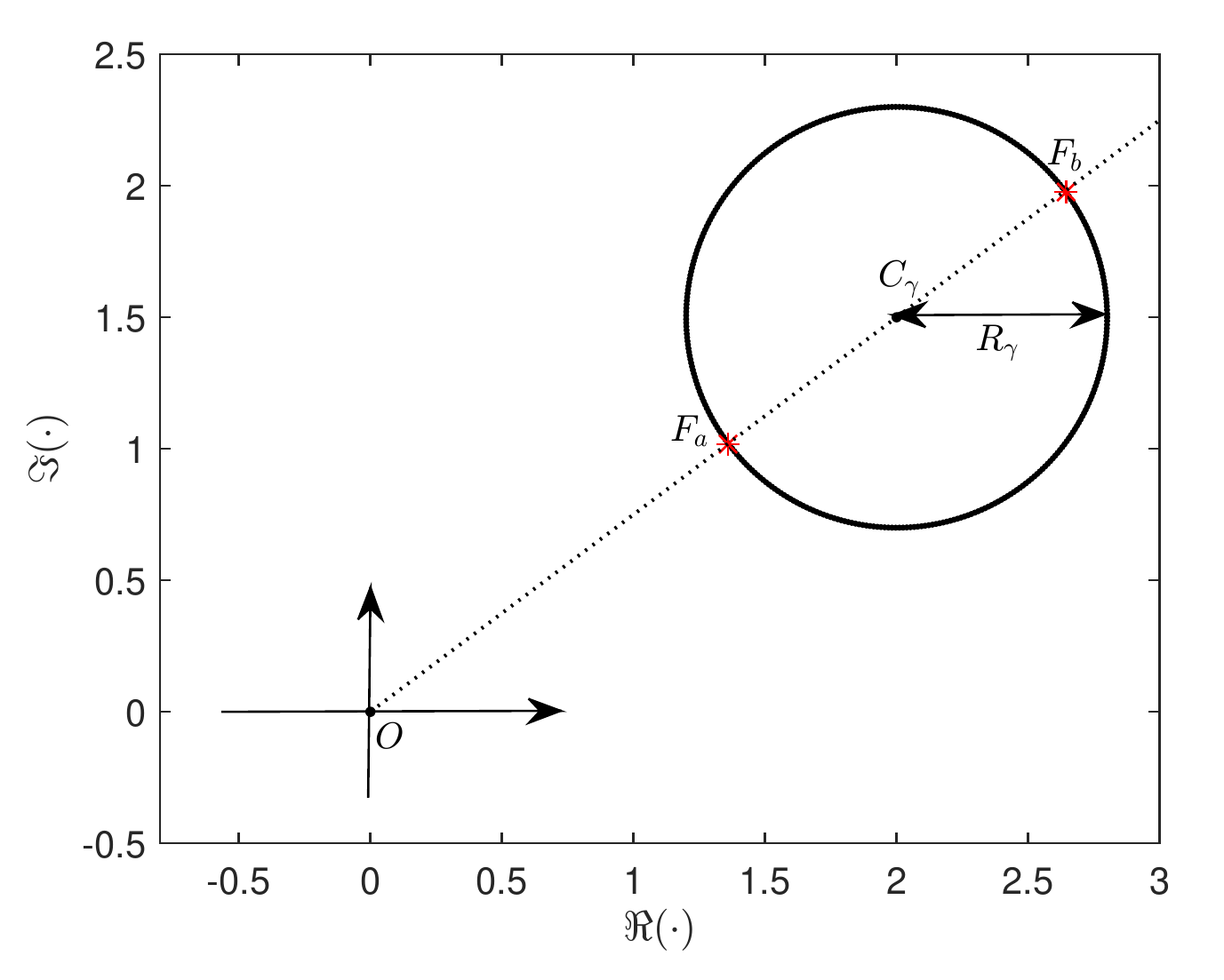}
	\caption{Geometric distribution of $ {\gamma}_k $.}
	\label{controlall}
\end{figure}

To have a better understanding, the locations of $ \gamma_{k,a} $ and $ \gamma_{k,b} $ have
been illustrated in Fig. \ref{controlall}.
Obviously, once the optimal $ {\gamma}_{k,\star} $ has been obtained, we can update the weight vector as
\begin{align}\label{modi120}
{\bf w}_k={\bf w}_{k-1}+{\gamma}_{k,\star}{\bf v}_{k}.
\end{align}
This completes the update of weight vector of the $ k $th step.

\subsection{Update of the Inversion of VCM}

Since the calculation of $\gamma_{k,\star}$ requires
the inversion of VCM (i.e., ${\bf T}^{-1}_{k-1} $)
which is assumed to be Hermitian in Proposition 2, in the sequel we shall discuss how to update $ {\bf T}^{-1}_{k} $
(in order to make the next step of response control
feasible) and how to guarantee the Hermitian property. To address these two problems, we first assume that
$ {\bf T}_{k-1} $ is Hermitian and the optimal $ {\gamma}_{k,\star} $ has been obtained with the aid of Proposition 2.
Then, from \eqref{modi0005} we have
\begin{align}\label{modi0006}
-{\beta_{k,\star}}/
\left({1+{\beta_{k,\star}}{\xi}_k}\right)=
{{\gamma}_{k,\star}}/{{\xi}_c}
\end{align}
where $ \beta_{k,\star}=\Psi^{-1}_{k}({\gamma}_{k,\star}) $ denotes the $ {\rm INR} $ corresponding to $ {\gamma}_{k,\star} $ in the $ k $th step.
Obviously, the combination of \eqref{modi0008} and \eqref{modi0006} yields
\begin{align}\label{modi0009}
{\bf T}^{-1}_{k}
={\bf T}^{-1}_{k-1}+
\dfrac{{\gamma}_{k,\star}{\bf v}_k{\bf v}^{\mH}_k}{\xi_c}.
\end{align}
Therefore, $ {\bf T}^{-1}_{k} $ can be calculated out straightforwardly (without the calculation of $ {\bf T}_{k} $ and its inversion), once the optimal $ {\gamma}_{k,\star} $ obtained.

To further explore the Hermitian property of $ {\bf T}_{k} $, let us revisit Proposition 2 and get
\begin{align}\label{modi0109}
\dfrac{{\gamma}_{k,\star}}{\xi_c}=-\dfrac{\left(\|{\bf c}_{\gamma}\|_2\pm R_{\gamma}\right)\chi}
{\|{\bf c}_{\gamma}\|_2{\bf H}_k(2,2)}\in\mathbb{R}
\end{align}
which is a real-valued number. Thus, it is known that $ {\bf T}_{k} $
is Hermitian as long as $ {\bf T}_{k-1} $ is Hermitian. Accordingly, it can be readily concluded that if we set $ {\bf T}_{0}={\bf I} $,
then $ {\bf T}_{i} $ is Hermitian for $ i=1,2,\cdots,k $.

Now, it is seen that the response levels can be successively adjusted by assigning virtual interferences.
Therefore, we can utilize the above update procedures iteratively to fulfill the prescribed response control requirement. Finally, the proposed OPARC method is summarized in Algorithm \ref{coded}.

\begin{algorithm}[!t]
	\caption{OPARC Algorithm}\label{coded}
	\begin{algorithmic}[1]
		\State give the initial weight vector $ {\bf w}_{0}={\bf a}(\theta_0) $
		and set $ {\bf T}_0={\bf I} $, prescribe the direction $ {\theta_{k}} $
		and the corresponding desired level
		$ {\rho_{k}} $, $ k=1,2,\cdots $
		\For{ $k = 1,2,\cdots, $}
		\State calculate $ {\gamma}_{k,\star} $ from \eqref{eqn0361} and
		obtain $ {\bf v}_k $ from \eqref{modi0021}
		\State update $ {\bf w}_k $ as $ {\bf w}_k={\bf w}_{k-1}+{\gamma}_{k,\star}{\bf v}_{k} $
		\State update $ {\bf T}^{-1}_{k} $ as $ {\bf T}^{-1}_{k}
		={\bf T}^{-1}_{k-1}+
		{{\gamma}_{k,\star}{\bf v}_k{\bf v}^{\mH}_k}\big/{\xi_c} $
		\EndFor
		\State output $ {\bf w}_k $	directly		
	\end{algorithmic}
\end{algorithm}

\section{Some Properties of $ \textrm{OPARC} $}
In the previous section, we have shown that the response level at a certain
direction can be optimally and flexibly adjusted by assigning a virtual interference.
Instead of determining $ \beta_k $ (i.e., the INR of the virtue interference assigned in the $ k $th step),
an alternative parameter $ \gamma_k $, mapping of $ \beta_k $, is chosen to facilitate the algorithm derivation.
However, since INR has a physical meaning and is always a non-negative value in a real data case, it is worth examining the direct
relationship between the response level and $ \beta_k $.
To do so, in this section we continue the analysis on the OPARC scheme, mainly
focus on the selection of INR $ \beta_k $ rather than its mapping $ \gamma_k $.

\subsection{Geometrical Distribution of $ {\beta_k} $}
As shown in \eqref{modi0005}, $ {\gamma}_k $ is a mapping of
$ \beta_k $, and $ \beta_k $ can be expressed with respect to  $ {\gamma}_k $ as
\begin{align}\label{modi0001}
{\beta}_k=-{\gamma_k}\big/
\left({{\xi}_c+{\gamma_k}{\xi_k}}\right)=\Psi^{-1}_{k}({\gamma_k})
\end{align}
where $ \Psi^{-1}_{k}(\cdot) $ is the inverse function of $ \Psi_{k}(\cdot) $ in \eqref{modi0005}. Thus, $ \beta_k $ can be calculated once $ {\gamma}_k $ is available. 
For the trajectory of $ \beta_k $ when the array response level $ \rho_k $ is satisfied at angle $ \theta_k $,
let us recall Proposition 1 and express $ {\gamma_k} $ as
\begin{align}\label{modi0051}
{\gamma_k}={\bf c}_{\gamma}(1)+j{\bf c}_{\gamma}(2)+R_{\gamma}e^{j\varphi}
\end{align}
where $ {\bf c}_{\gamma} $ and $ R_{\gamma} $ are the center and the radius of the circle
given in Proposition 1, $ \varphi $ can be any real-valued number.
Substituting \eqref{modi0051}
into \eqref{modi0001}, one gets
\begin{align}\label{modi0052}
{\beta}_k=\left({p_1+p_2e^{j\varphi}}\right)/\left({q_1+q_2e^{j\varphi}}\right)
\end{align}
where $ p_l $ and $ q_l $ ($ l=1,2 $) are complex numbers satisfying
\begin{subequations}\label{modi0056}
	\begin{align}
	p_1&=-{\bf c}_{\gamma}(1)-j{\bf c}_{\gamma}(2),
	~~~~~~~~~~~p_2=-R_{\gamma}\\
	q_1&={\xi}_c+\left({\bf c}_{\gamma}(1)+j{\bf c}_{\gamma}(2)\right){\xi}_k,
	~~q_2=R_{\gamma}{\xi}_k.
	\end{align}
\end{subequations}

With some calculation, it is not difficult to have the following proposition.
\begin{theorem}
	The trajectory of $[{\Re}(\beta_k)~~{\Im}(\beta_k)]^{\mT} $ with $ \beta_k $
	satisfying \eqref{modi0052} is a circle $ {\mathbb{C}}_{\beta} $ with the center
	\begin{align}\label{eqn0513}
	{\bf c}_{\beta}=\left[
	{{\xi}_0}/({|{\xi}_c|^2-{{\xi}_0}{{\xi}_k}}),~0
	\right]^{\mT}
	\end{align}
	and the radius
	\begin{align}\label{eqn053}
	{R_{\beta}}={|{\xi}_c|}\big/\left[\sqrt{\rho_k}\cdot{\big||{\xi}_c|^2-{\xi}_0{\xi}_k\big|}\right]
	\end{align}
	i.e.,
	\begin{align}
	\beta_k={\bf c}_{\beta}(1)+j{\bf c}_{\beta}(2)+R_{\beta}e^{j\phi}
	\end{align}
where $ \phi $ can be any real-valued number.
\end{theorem}

Similar to $ \gamma_k $, all the $ \beta_k $ on the above circle can be used to
precisely
adjust the response level at $ \theta_k $ to its desired level $ \rho_k $.
An interesting difference with $ \gamma_k $ is that the calculations of $ {\bf c}_{\beta} $
and $ R_{\beta} $ do not require the knowledge of ${\bf w}_{k-1}$.
This implies that all $ \beta_k $'s (including the optimal one later) can be obtained without knowing any weight vectors. On the contrary, the determination of $\gamma_k$ relies on the availability of the previous weight vector ${\bf w}_{k-1}$.

In addition, Proposition 3 implies that the center of
the trajectory of $[{\Re}(\beta_k)~~{\Im}(\beta_k)]^{\mT}  $ is located on the real axis,
and is independent of the desired level $ {\rho_k} $.

\subsection{Determination of the Optimal $ {\beta_{k}} $}
Among all the valid $\beta_k$ for the control of the array response level $\rho_k$ at $\theta_k$, 
the optimal one is the one that maximizes the array gain.
Therefore, the following constrained optimization problem can be formulated to select the optimal 
$ {\beta_{k}} $:
\begin{subequations}\label{p2qu011}	
	\begin{align}
	\mathop {{\rm{maximize}}}\limits_{{\beta _k}} &~~~G_k=
	{|{\bf w}^{\mH}_k{\bf a}(\theta_0)|^2}/{|{\bf w}^{\mH}_k{\bf T}_k{\bf w}_k|}\\
	\label{qu211}{\rm subject~to}&~~~L(\theta_k,\theta_0)={\rho_k}\\
	\label{qu212}&~~~{\bf w}_k={\bf w}_{k-1}+\Psi_{k}({\beta_k}){\bf T}^{-1}_{k-1}{\bf a}(\theta_k)
	\end{align}
\end{subequations}
where the parameter $ \gamma_k $ has been replaced in \eqref{qu212} by $ \Psi_{k}({\beta_k}) $.
Clearly, the above optimization problem \eqref{p2qu011} is equivalent to problem \eqref{qu011}.
Therefore, the optimal solution (denoted as $ {\beta}_{k,\star} $) of \eqref{p2qu011} can be readily obtained
by utilizing the mapping as
\begin{align}\label{mo0001}
{\beta}_{k,\star}=\Psi^{-1}_{k}({\gamma_{k,\star}}).
\end{align}
Combining the result of $ {\gamma}_{k,\star} $ in \eqref{eqn0361} with some calculation, we can derive Proposition 4 below.

\begin{theorem}
	The optimal solution of \eqref{p2qu011} is given by
	\begin{align}\label{qua011}
	{\beta}_{k,\star}=\left\{
	\begin{array}{cc}
	{\beta}_{k,r},&{\textrm{if}}~~~{-1}/{\xi_k}>{{\xi}_0}/\left({|{\xi}_c|^2-{{\xi}_0}{{\xi}_k}}\right)\\
	{\beta}_{k,l},&{\textrm{otherwise}}
	\end{array} \right.
	\end{align}
	where $ \beta_{k,r} $ and $ \beta_{k,l} $ are the intersections of circle $ {\mathbb{C}}_{\beta} $
	and the real axis $ {\Im}(\cdot)=0 $:
	\begin{align}
	\label{qua015}{\beta}_{k,r}&=R_{\beta}+{{\xi}_0}/\left({|{\xi}_c|^2-{{\xi}_0}{{\xi}_k}}\right)\\
	\label{qua016}{\beta}_{k,l}&=-R_{\beta}+{{\xi}_0}/\left({|{\xi}_c|^2-{{\xi}_0}{{\xi}_k}}\right).
	\end{align}	
\end{theorem}

\begin{algorithm}[!t]
	\caption{OPARC Algorithm (an Equivalent Variant)}\label{coded2}
	\begin{algorithmic}[1]
		\State give $ {\bf a}(\theta_0) $ and set $ {\bf T}_0={\bf I} $, specify the direction $ {\theta_{k}} $
		and the corresponding desired level
		$ {\rho_{k}} $ with $ k=1,2,\cdots $
		\For{ $k = 1,2,\cdots, $}
		\State calculate $ {\beta}_{k,\star} $ from \eqref{qua011}
		\State update $ {\bf T}_k $ as $ {\bf T}_k={\bf T}_{k-1}+{\beta}_{k,\star}{\bf a}(\theta_k){\bf a}^{\mH}(\theta_k) $
		\EndFor
		\State calculate $ {\bf w}_k={\bf T}^{-1}_k{\bf a}(\theta_0) $			
	\end{algorithmic}
\end{algorithm}

It is not hard to see from \eqref{qua011} that the optimal $ {\beta}_{k,\star}$ is a real-valued number,
while the valid $ \beta_k $ in Proposition 3 for the array response control may be complex valued. 
However, as mentioned earlier, the physical meaning of $ \beta_k $ is the INR as it is used in \eqref{eqn1004} and it cannot be 
negative.  From \eqref{qua011}-\eqref{qua016}, the solved optimal $ \beta_{k,\star} $ may be negative, which might be because there is no assumption of the used VCM 
$ {\bf T}_{k-1} $ being non-negative definite. This will be studied together with the update of the VCM below.
On the other hand, if $ {\bf T}_{k-1} $ is Hermitian, then $ {\bf T}_k={\bf T}_{k-1}+{\beta}_{k,\star}{\bf a}(\theta_k){\bf a}^{\mH}(\theta_k) $ is also 
Hermitian, since $ \beta_{k,\star} $ is real.
This is consistent with the inference obtained in the paragraph below Eqn. \eqref{modi0109}.
Finally, it is obvious that the optimal $ {\beta}_{k} $ in \eqref{qua011} does not depend on the knowledge of the weight vectors in the previous steps.

Once the optimal $ \beta_{k} $ has been obtained, we can express the VCM at the current stage as
\begin{align}\label{coroparc}
{\bf T}_k={\bf T}_{k-1}+{\beta}_{k,\star}{\bf a}(\theta_k){\bf a}^{\mH}(\theta_k)=
{\bf I}+{\bf A}_k
{\bf \Sigma}_{k}{\bf A}^{\mH}_k
\end{align}
where $ {\bf A}_{k} \triangleq[{\bf a}(\theta_1),\cdots,{\bf a}(\theta_{k})] $  and $ {\bf \Sigma}_{k} $ is a diagonal matrix containing all $ \beta $'s of virtual interferences, i.e.,
\begin{align}\label{qua520}
{\bf \Sigma}_{k}={\rm Diag}
\left(\left[{\beta}_{1,\star},{\beta}_{2,\star},\cdots,{\beta}_{k,\star}\right]\right).
\end{align}
Accordingly, we have $ {\bf w}_k={\bf T}^{-1}_k{\bf a}(\theta_0) $.
To make it clear, the variant of the OPARC method is summarized in Algorithm \ref{coded2}.
Note that the calculation of intermediate weight vectors is avoided,
due to the fact that neither the calculation of $ \beta_{k,\star} $ nor
$ {\bf T}_k $ relies on weight vectors.
Therefore, the procedure of array response control is simplified.

\begin{figure}[!t]
	\centering
	\includegraphics[width=3.0in]{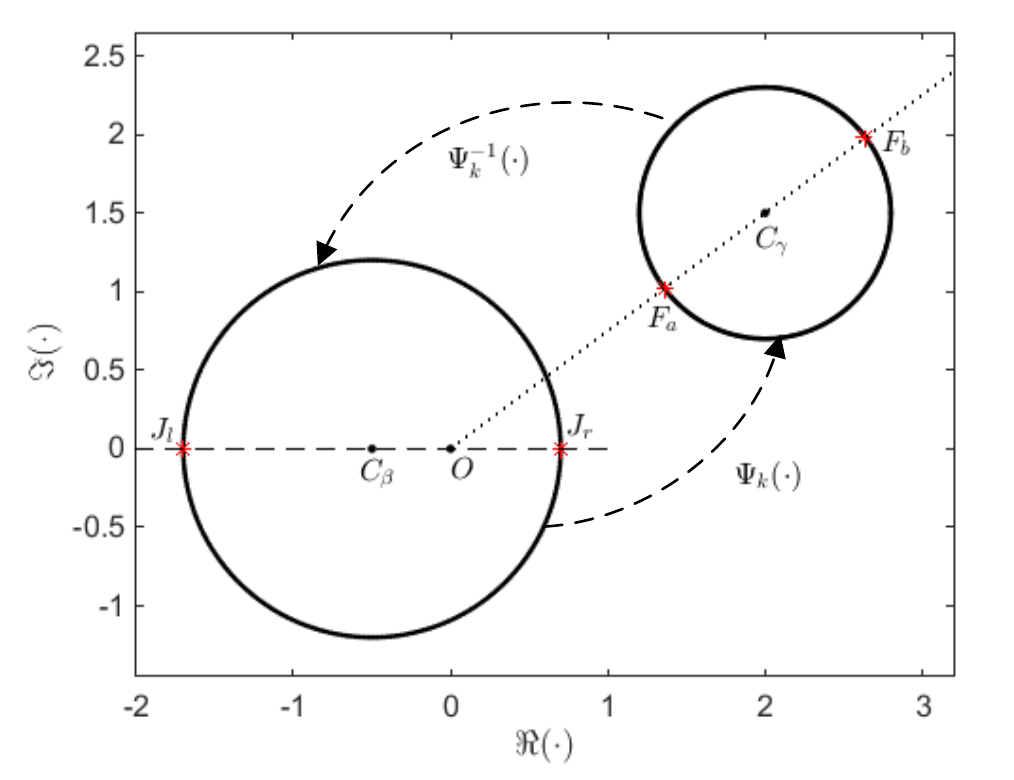}
	\caption{Illustration of the mapping $ \Psi_k(\cdot) $.}
	\label{papermapping}
\end{figure}

Before proceeding, it is interesting to provide a deep insight and a geometrical perspective on
the relationship between $ \gamma_k $ and $ \beta_k $.
It is not hard to see that the condition $ {-1}/{\xi_k}>{{\xi}_0}/\left({|{\xi}_c|^2-{{\xi}_0}{{\xi}_k}}\right) $ in
	\eqref{qua011} is equivalent to the condition $ \zeta>0 $ in \eqref{eqn0361}, if and only if
	$ {\rho_k}{\xi_0}<{\xi_k} $.
It implies that the conditions for selecting $ {\beta}_{k,\star} $ between $ {\beta}_{k,l} $ and $ {\beta}_{k,r} $ and selecting $ {\gamma}_{k,\star} $ between $ {\gamma}_{k,a} $ and $ {\gamma}_{k,b} $ 
may be different and are the same under a certain condition, i.e., $ {\rho_k}{\xi_0}<{\xi_k} $. Thus, we have
\begin{align}\label{qua311}
\left\{
\begin{array}{cc}
{\gamma_{k,a}}=\Psi_k({\beta}_{k,r}),{\gamma_{k,b}}=\Psi_k({\beta}_{k,l}),&{\textrm{if}}~{\rho_k}{\xi_0}<{\xi_k}\\
{\gamma_{k,a}}=\Psi_k({\beta}_{k,l}),{\gamma_{k,b}}=\Psi_k({\beta}_{k,r}),&{\textrm{otherwise}}.
\end{array} \right.
\end{align}
To have an intuitive perspective on $ \Psi_k(\cdot) $,
a geometrical illustration is given in Fig. \ref{papermapping},
where $ J_r $, $ J_l $, $ F_a $ and $ F_b $ stand for the points
$ [{\Re(\beta_{k,r})}~{\Im(\beta_{k,r})}]^{\mT} $, $ [{\Re(\beta_{k,l})}~{\Im(\beta_{k,l})}]^{\mT} $,
$ [{\Re(\gamma_{k,a})}~{\Im(\gamma_{k,a})}]^{\mT} $
and $ [{\Re(\gamma_{k,b})}~{\Im(\gamma_{k,b})}]^{\mT} $, respectively.

\subsection{Positive Definite Virtual Covariance Matrices}

Firstly, the following conclusion, which simplifies the selection of $ \beta_{k,\star} $, can be obtained.
\begin{theorem}
	If $ {\bf T}_{k-1}\in\mathbb{S}^N_{++} $, we have
	\begin{align}\label{qua017}
	{\beta}_{k,\star}={\beta}_{k,r}
	=({|\xi_c|-\sqrt{{\rho}_k}{\xi}_0})/[{\sqrt{{\rho}_k}({{\xi}_0}{{\xi}_k}-|{\xi}_c|^2)}].
	\end{align}
	Furthermore, if $ {\bf T}_{k-1}\in\mathbb{S}^N_{++} $,
	then $ {\bf T}_{k}\in\mathbb{S}^N_{++} $ if and only if
	$ {\rho}_k<{\xi}^2_k/|{\xi}_c|^2 $.
\end{theorem}
\begin{IEEEproof}
	See Appendix B.
\end{IEEEproof}

%
Similar to the argument in the paragraph below Eqn. \eqref{barrho}, 
$ {\xi}^2_k/|{\xi}_c|^2 $ is in general greater than 1 and it is assumed $ \rho_k\leq1 $. Thus, we have $ {\rho}_k<{\xi}^2_k/|{\xi}_c|^2 $.
As a consequence, in each step of weight vector update, we have $ {\bf T}_{k}\in\mathbb{S}^N_{++} $ and $ {\beta_{k,\star}}={\beta}_{k,r} $,
as long as $ {\bf T}_{k-1}\in\mathbb{S}^N_{++} $.
Since in our algorithm $ {\bf T}_{0}={\bf I} $ is taken as the initial VCM,
we have $ {\bf T}_{k}\in\mathbb{S}^N_{++} $ and $ {\beta}_{k,\star}={\beta}_{k,r} $.

%

\begin{theorem}
	If $ {\bf T}_{k-1}\in\mathbb{S}^N_{++} $, then
	\begin{subequations}\label{eqn56}
	\begin{align}
\beta_{k,\star}\geq 0&\Leftrightarrow{|\xi_c|\geq\sqrt{{\rho}_k}{\xi}_0}\\
\beta_{k,\star}< 0&\Leftrightarrow{|\xi_c|<\sqrt{{\rho}_k}{\xi}_0}.
\end{align}	
\end{subequations}
\end{theorem}
\begin{IEEEproof}
	From \eqref{eqn2067} in the proof of Proposition 5 in Appendix B, we have 
	$ {{\xi}_0}{{\xi}_k}-|{\xi}_c|^2>0 $ provided that $ {\bf T}_{k-1}\in\mathbb{S}^N_{++} $.
	Then, from \eqref{qua017}, the proof of \eqref{eqn56} is completed.
\end{IEEEproof}

Substituting the definitions of $ \xi_c $ and $ \xi_0 $ into \eqref{eqn56} and using $ {\bf w}_{k-1}={\bf T}^{-1}_{k-1}{\bf a}(\theta_0) $,
we have
\begin{subequations}\label{eqn57}
	\begin{align}
	\beta_{k,\star}\geq 0&\Leftrightarrow{\rho}_k\leq
	{\big|
		{\bf w}^{\mH}_{k-1}{\bf a}(\theta_k)
		\big|^2}/
	{\big|
		{\bf w}^{\mH}_{k-1}{\bf a}(\theta_0)
		\big|^2}\\
	\beta_{k,\star}< 0&\Leftrightarrow{\rho}_k>
	{\big|
		{\bf w}^{\mH}_{k-1}{\bf a}(\theta_k)
		\big|^2}/
	{\big|
		{\bf w}^{\mH}_{k-1}{\bf a}(\theta_0)
		\big|^2}.
	\end{align}	
\end{subequations}
Notice that $ {\big|
		{\bf w}^{\mH}_{k-1}{\bf a}(\theta_k)
		\big|^2}/
	{\big|
		{\bf w}^{\mH}_{k-1}{\bf a}(\theta_0)
		\big|^2} $ above represents the normalized response at $ \theta_k $, of the previous weight vector $ {\bf w}_{k-1} $.
Clearly, \eqref{eqn57} shows that the resultant $ \beta_{k,\star} $ is non-negative if the desired level $ \rho_k $
is lower than the response level at $ \theta_k $ of the previous weight vector $ {\bf w}_{k-1} $. Otherwise, a negative $ \beta_{k,\star} $
is obtained if it is required to elevate the previous response level of $ \theta_k $.
We can see that the negative $ \beta_{k,\star} $ is still meaningful in our discussion of array response control
using virtual interferences,
although it cannot occur in a real data covariance matrix with real interferences.

In addition to the above two propositions, the following result can be obtained.

\begin{theorem}
	If $ {\bf T}_{k-1}\in\mathbb{S}^N_{++} $, then problem \eqref{p2qu011} has the same optimal solution as that of the following one
	\begin{subequations}\label{qu552}	
		\begin{align}
		\mathop {{\rm{maximize}}}\limits_{{\beta _k}}&~~~\dfrac{|{\bf w}^{\mH}_k{\bf a}(\theta_0)|^2}
		{{\bf w}^{\mH}_{k}{\bf T}_{k-1}{\bf w}_{k}}\\
\label{qu52}{\rm subject~to}&~~~L(\theta_k,\theta_0)={\rho_k}\\
		&~~~{\bf w}_k={\bf w}_{k-1}+\Psi_k({\beta_k}){\bf v}_{k}.
		\end{align}
	\end{subequations}
\end{theorem}
\begin{IEEEproof}
	See Appendix C.
\end{IEEEproof}

Interestingly, from Proposition 7, it is known that under the same
constraints (i.e., \eqref{qu211} and \eqref{qu212}), the optimal $ {\beta_{k}} $ to \eqref{p2qu011} also maximizes
the previous array gain, in which only $ {\bf T}_{k-1} $ (but not $ {\bf T}_{k} $) is taken into consideration.

For instance, consider the case when $ {\bf T}_{k-1} $ is a real normalized noise-plus-interference covariance matrix
(i.e., $ {\bf T}_{k-1} $ is calculated from real data that contains both noise and interference),
and one applies the OPARC scheme to realize a specific array response control task in \eqref{qu52}
by assigning a virtual interference at $ \theta_k $.
Then, it is seen from Proposition 7 that the optimal $ \beta_{k,\star} $ of problem \eqref{p2qu011}
also maximizes the real output SINR (not taking the virtual interference into consideration) 
of beamformer.
This property will be further exploited in the companion paper \cite{p3} to design
an adaptive beamformer with specific constraint.


\section{Comparison with ${\textrm A}^2{\textrm{RC}}$}
In the above sections, the optimal values of $ \gamma_k $ and $\beta_k$ of the virtual interference assigned
in the $ k $th step are specified.
Meanwhile, useful conclusions are obtained and two versions of OPARC are described.
In this section, comparisons will be carried out to elaborate the differences
between the recent ${\textrm A}^2{\textrm{RC}}$ algorithm \cite{snrf41} and the above OPARC algorithm from two perspectives.

\subsection{Comparison on the Formula Updating}

In the ${\textrm A}^2{\textrm{RC}}$ method, the weight vector is updated as
\begin{align}\label{qua7511}
{{\bf w}}_{k}&={\bf w}_{k-1}+\mu_k{\bf a}(\theta_k)
\end{align}
where $ \mu_k $ is the hyperparameter to be optimized.
To minimize the deviation between the resultant responses of adjacent two steps, and meanwhile, 
avoid the computationally inefficient global search, $ {\mu}_{k} $ is empirically selected in \cite{snrf41}
as ${\mu_{k,a}} $, which is the solution to the following problem:
\begin{subequations}\label{e2156}
\begin{align}
\mathop {{\rm{minimize}}}\limits_{{\mu _k}}&~~~|\mu_k|\\
{\rm subject~to}&~~~
\begin{bmatrix}	{\Re}(\mu_k)& {\Im}(\mu_k)\end{bmatrix}^{\mT}\in\mathbb{C}_{\mu}
\end{align}
\end{subequations}
where $ \mathbb{C}_{\mu} $ is the following circle:
\begin{align}
\mathbb{C}_{\mu}=\left\{
[	{\Re}(\mu_k)~~ {\Im}(\mu_k)]^{\mT}\Big|
\big\|[	{\Re}(\mu_k)~~ {\Im}(\mu_k)]^{\mT}-{\bf c}_{\mu}\big\|_2=R_{\mu}
\right\}\nonumber
\end{align}	
with the center
\begin{align}\label{e2qn0030}
{\bf c}_{\mu}=\dfrac{1}{{\bf Q}_k(2,2)}
\begin{bmatrix}-{\Re}\left[{\bf Q}_k(1,2)\right]\\ {\Im}\left[{\bf Q}_k(1,2)\right]\end{bmatrix}
\end{align}
and the radius
\begin{align}\label{eqn20031}
R_{\mu}={\sqrt{-{\rm det}({\bf Q}_k)}}\big/{|{\bf Q}_k(2,2)|}
\end{align}
where the matrix $ {\bf Q}_k $ satisfies
\begin{align}
{\bf Q}_k\!=\![{\bf w}_{k\!-\!1}~~{\bf a}(\!\theta_k\!)]^{\mH} \!	\left(\! 	{\bf a}(\!\theta_k\!) {\bf a}^{\mH}(\!\theta_k\!)\!\!-\!\!\rho_k{\bf a}(\!\theta_0\!){\bf a}^{\mH}(\!\theta_0\!)\!\right)\! [	{\bf w}_{k\!-\!1}\!~~{\bf a}(\!\theta_k\!)].\nonumber
\end{align}
Note that such an empirical selection may not perform well under all circumstances.
As a matter of fact, this scheme may even lead to severe pattern distortion, as we will
show later in simulations in Section V.

In the OPARC algorithm, the weight vector is updated via \eqref{qu5212}.
It is seen that, different from the ${\textrm A}^2{\textrm{RC}}$ algorithm, a scaling of $ {\bf T}^{-1}_{k-1}{\bf a}(\theta_k) $ is added to $ {\bf w}_{k-1} $, 
and $ {\gamma}_k{\bf T}^{-1}_{k-1}{\bf a}(\theta_k) $ makes the resultant
$ {\bf w}_{k} $ be an optimal weight vector.

Additionally, in the proposed OPARC algorithm, we optimize the parameter $ \gamma_k $ by
	maximizing the array gain when the preassigned  response level is satisfied.

To have a similar weight form with ${\textrm A}^2{\textrm{RC}}$, it is shown in Appendix D that we
can reformulate \eqref{qu5212} as
\begin{align}\label{qua1}
{\bf w}_k&={\bf w}_{k-1}+
{\gamma}_k{\bf A}(\theta_k,\cdots,\theta_{1}){\bf d}_k\nonumber\\
&={\bf w}_{k-1}+
{\gamma}_k{\bf a}(\theta_k)+{\gamma}_k{\bf A}(\theta_{k-1},\cdots,\theta_{1})\bar{\bf d}_k
\end{align}
where $ {\bf A}(\theta_k,\cdots,\theta_{1})\triangleq[{\bf a}(\theta_k),\cdots,{\bf a}(\theta_{1})] $,
with $ \theta_i $ ($ 1\leq i\leq k-1 $) denoting the angles of interferences that assigned previously,
$ {\bf d}_k $ is a $ k\times1 $ vector with its first element 1,
$ \bar{\bf d}_k $ is a $ (k-1)\times1 $ vector obtained by removing the first element from $ {\bf d}_k $.
From \eqref{qua1}, it is observed that the added component to the previous weight vector $ {\bf w}_{k-1} $ in 
	$ {\bf w}_k $ is a linear combination of the steering vectors of all interferences 
(including both the current $ {\bf a}(\theta_k) $ and the previous $ {\bf a}(\theta_1),\cdots,{\bf a}(\theta_{k-1}) $). 
On the contrary, in the ${\textrm A}^2{\textrm{RC}}$ algorithm, the added component is a scaling of the steering vector of the single interference to be 
assigned (i.e., $ {\bf a}(\theta_k) $).
Furthermore, we can obtain the following corollary of Proposition 2, which describes
a similarity between ${\textrm A}^2{\textrm{RC}}$ and OPARC.

\newtheorem{theorem2}{Corollary}
\begin{theorem2}
	In the first step of weight update (i.e, $ {\bf w}_0={\bf a}(\theta_0) $, $ {\bf T}_0={\bf I} $),
	if $ {\rho_1}\leq
	\|{\bf a}(\theta_1)\|^2_2/\|{\bf a}(\theta_0)\|^2_2 $, then $ {\mu}_{1,\star}={\gamma}_{1,\star} $,
	otherwise, $ {\mu}_{1,\star}={\gamma}_{1,\times} $,
	where
${\gamma}_{1,\times}=\left\{{\gamma_{1,a}},{\gamma}_{1,b}\right\}\setminus{\gamma}_{1,\star}$.
\end{theorem2}
\begin{IEEEproof}
	See Appendix E.	
\end{IEEEproof}

From Corollary 1, it is known that in the first step of the weight vector update, ${\textrm A}^2{\textrm{RC}}$
will lead to the same result as OPARC, provided that $ {\rho_1}\leq
\|{\bf a}(\theta_1)\|^2_2/\|{\bf a}(\theta_0)\|^2_2 $.
Otherwise, the inferior parameter $ \gamma_{1,\times} $ (in the sense of array gain)
will be adopted by ${\textrm A}^2{\textrm{RC}}$.

\subsection{Comparison on INRs of Virtual Interferences}
We next compare the INRs of virtual interferences
to show an essential difference between
${\textrm A}^2{\textrm{RC}}$ and OPARC. To begin with, we express the weight vector of ${\textrm A}^2{\textrm{RC}}$ in the $ k $th step as
\begin{align}\label{qua511}
{{\bf w}}_{k}&={\bf a}(\theta_0)+
\mu_1{\bf a}(\theta_1)+\cdots+\mu_k{\bf a}(\theta_k)\nonumber\\
&={\bf a}(\theta_0)+{\bf A}_{k}{\bf b}_{k}
\end{align}
where $ {\bf b}_{k}\triangleq[\mu_1,\mu_2,\cdots,\mu_k]^{\mT} $.
Note from the above that the update of the weight vector $ {\bf w}_k $ in ${\textrm A}^2{\textrm{RC}}$ does not 
depend on any VCM and no VCM update is needed. However, in order to compare the INRs, 
we need to associate it to a VCM that may be implicit/virtual. To do so, we rewrite the weight vector as
\begin{align}\label{qua513}
{{\bf w}}_{k}&={\breve{\bf T}}^{-1}_k{\bf a}(\theta_0)\nonumber\\
&={\bf a}(\theta_0)-{\bf A}_{k}\left({\bf I}+{\breve{\bf \Sigma}}_k{\bf A}^{\mH}_{k}{\bf A}_{k}\right)^{-1}
{\breve{\bf \Sigma}}_k{\bf A}^{\mH}_{k}{\bf a}(\theta_0)
\end{align}
where
$ {\breve{\bf T}}_k={\bf I}+{\bf A}_k{\breve{\bf \Sigma}}_k{\bf A}^{\mH}_k $
denotes a VCM,
$ {\breve{\bf \Sigma}}_k={\rm Diag}([\breve{\beta}_{k,1},\breve{\beta}_{k,2},\cdots,\breve{\beta}_{k,k}]) $
specifies the INR of the interference at 
$ \theta_i $ ($i=1,\cdots, k $) when completing
the current $ k $th step of the weight vector update.

Note that no interference was assigned at the current $ {\theta_k} $ in the previous $ k-1 $ steps of the response control.
It is shown from Appendix F that in the $ k $th step of the weight update
	of the ${\textrm A}^2{\textrm{RC}}$ algorithm, the INR of the virtual interference assigned at $ \theta_k $ is
\begin{align}\label{prooff1}
	{\breve{\beta}_{k,k}}=-\dfrac{{\mu}_k}{{\bf a}^{\mH}(\theta_k){\breve{\bf w}}_{k-1}+
		{\mu}_k\|{\bf a}(\theta_k)\|^2_2}.
\end{align}
In addition, $ k-1 $ new interferences are additively assigned at directions $ \theta_1,\cdots,\theta_{k-1} $
of ${\textrm A}^2{\textrm{RC}}$ to the previous $ (k-1) $th step.
Denote the INRs of these new interferences assigned at $ {\theta_i} $ 
in the $ k $th step of the weight update
as $ {\breve \Delta}_{k,i} $ ($ 1\leq i\leq k-1 $). Clearly they satisfy
\begin{align}\label{prff2}
{\breve \Delta}_{k,i}={\breve{\beta}_{k,i}}-{\breve{\beta}_{k-1,i}}.
\end{align}
It can be further derived (see Appendix F)
that
\begin{align}\label{prooff2}
{\breve \Delta}_{k,i}=\dfrac{{\mu}_k{\bf a}^{\mH}(\theta_i){\bf a}(\theta_k){\breve{\beta}^2_{k-1,i}}}
{{\mu}_i-{\mu}_k{\bf a}^{\mH}(\theta_i){\bf a}(\theta_k){\breve{\beta}_{k-1,i}}}.
\end{align}

Generally speaking, in the $ k $th step of ${\textrm A}^2{\textrm{RC}}$, the INRs of the newly-assigned interferences
(including both $ {\breve{\beta}_{k,k}} $ and $ {\breve \Delta}_{k,i} $ ($ 1\leq i\leq k-1 $)) are complex-valued numbers.
This is a difference between ${\textrm A}^2{\textrm{RC}}$ and OPARC.
Moreover, the above analysis shows that there are $ k-1 $ additional interferences assigned to the previously
controlled angles (i.e., $ \theta_1,\cdots,\theta_{k-1} $) in 
the $ k $th step of ${\textrm A}^2{\textrm{RC}}$, 
while in OPARC only a single interference is assigned (at $ \theta_k $).
Since our aim is to control the array response level at $ \theta_k $, the newly-assigned virtual interferences at 
$ \theta_1,\cdots,\theta_{k-1} $ actually bring undesirable array response variations at these adjusted angles.
According to the above notations, the VCM of ${\textrm A}^2{\textrm{RC}}$ satisfies implicitly:
\begin{align}\label{vsme}
{\breve{\bf T}}_k={\breve{\bf T}}_{k-1}+
{\bf A}_k{\rm Diag}([{\breve \Delta}_{k,1},\cdots,{\breve \Delta}_{k,k-1},\breve{\beta}_{k,k}])
{\bf A}^{\mH}_k
\end{align}
which is different from that of $ \textrm{OPARC} $ in \eqref{coroparc}.

To summarize, the main differences between the proposed OPARC algorithm and the existing
${\textrm A}^2{\textrm{RC}}$ algorithm include:
\begin{itemize}
	\item Different formulas of the weight update are employed in OPARC and ${\textrm A}^2{\textrm{RC}}$. 
	\item The resultant weight of OPARC can be guaranteed to be an optimal beamformer, while
	${\textrm A}^2{\textrm{RC}}$ method does not.
	\item The array gain is introduced to the parameter optimization of OPARC, while ${\textrm A}^2{\textrm{RC}}$ is not.
	\item The update of VCM is necessary for OPARC, while ${\textrm A}^2{\textrm{RC}}$
	 is free of this procedure although its VCM is implicitly updated by \eqref{vsme}.
	\item Two different strategies of virtual interference assigning are adopted
	in these two approaches. The INRs of OPARC are always real, but they may not be in ${\textrm A}^2{\textrm{RC}}$.
\end{itemize}

\begin{figure*}[!t]
	\centering
	\subfloat[Comparison of synthesized patterns at the 1st step]
	{\includegraphics[width=3.55in]{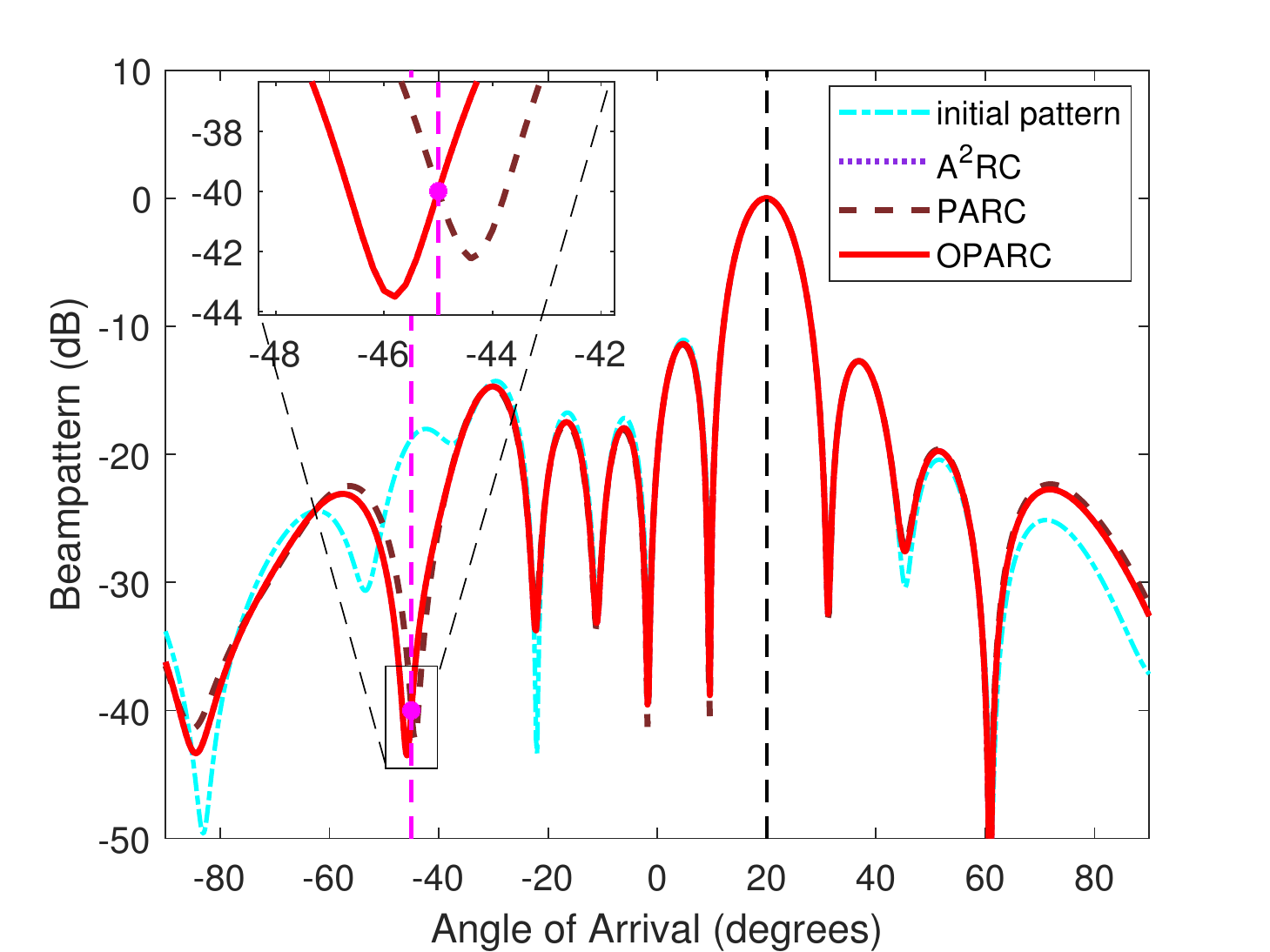}%
		\label{twosidestep1}}
	\hfil
	\subfloat[Comparison of synthesized patterns at the 2nd step]
	{\includegraphics[width=3.55in]{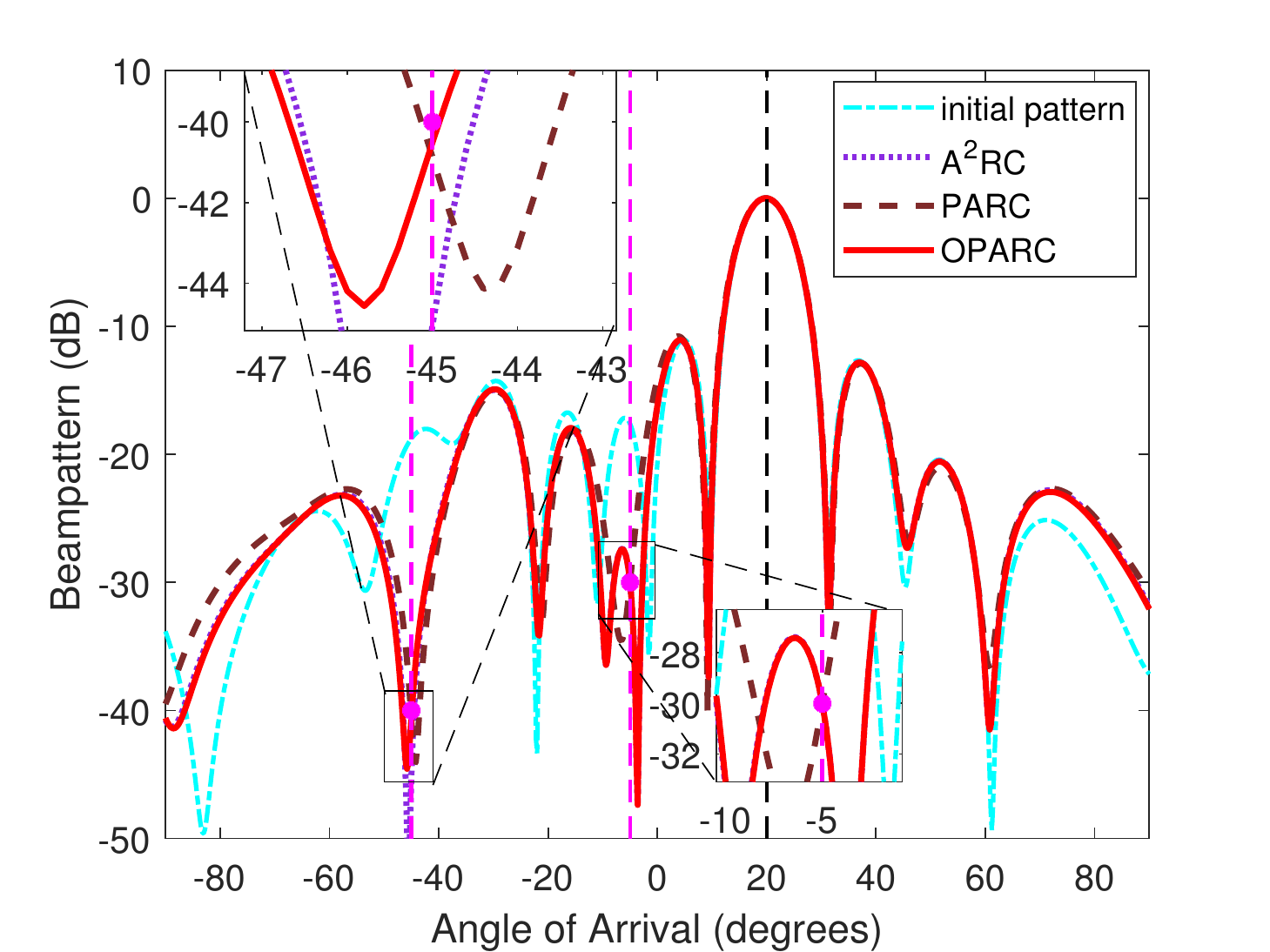}%
		\label{twosidestep2}}				
	\caption{Resultant pattern comparison (the first example).}
	\label{trans00182dqiwang4}
\end{figure*}
\begin{figure*}[!t]
	\centering
	\subfloat[Curves of $ D $ versus $ {\rho}_2 $]
	{\includegraphics[width=3.55in]{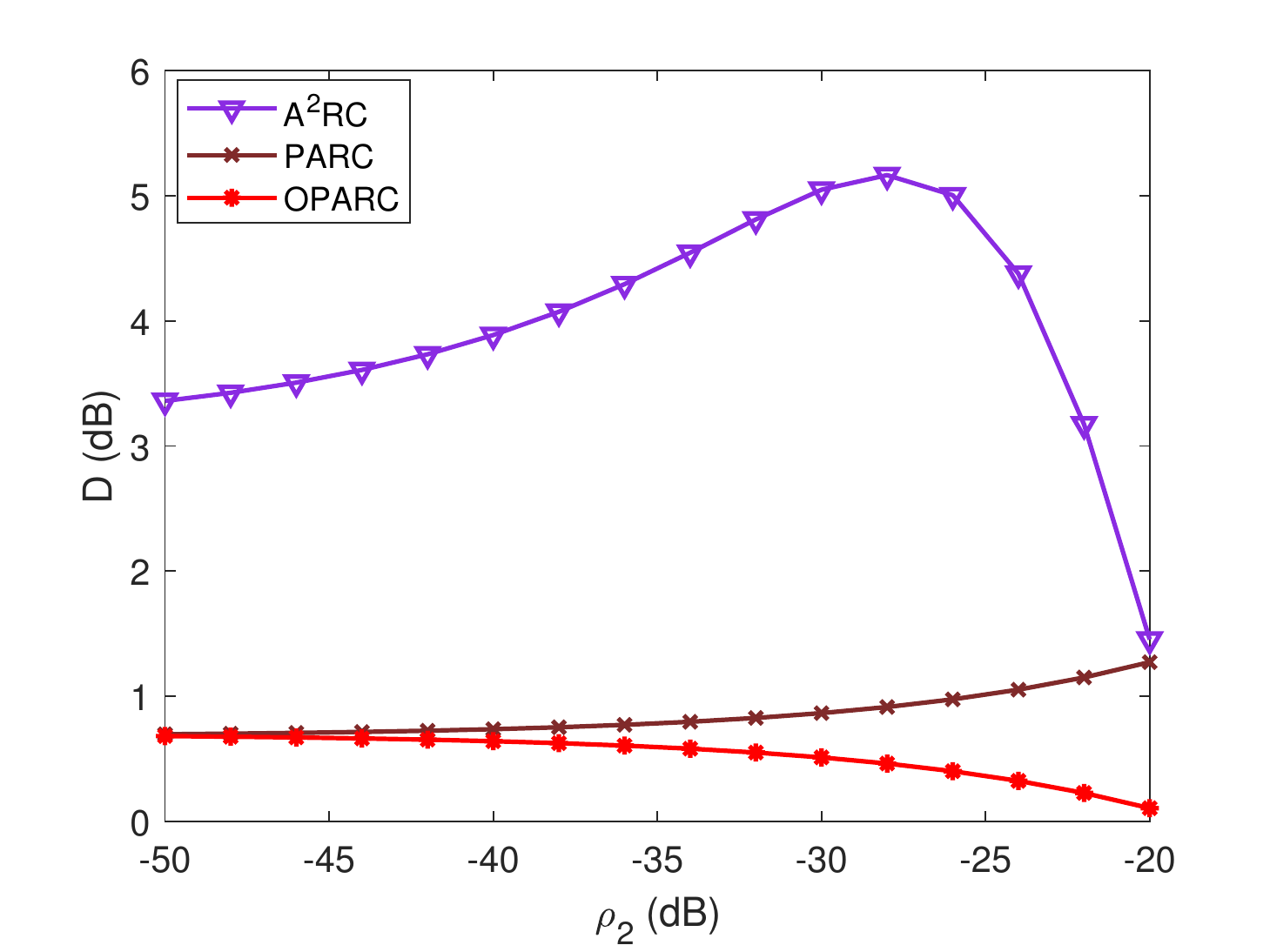}%
		\label{sideD}}
	\hfil
	\subfloat[Curves of $ J $ versus $ {\rho}_2 $]
	{\includegraphics[width=3.55in]{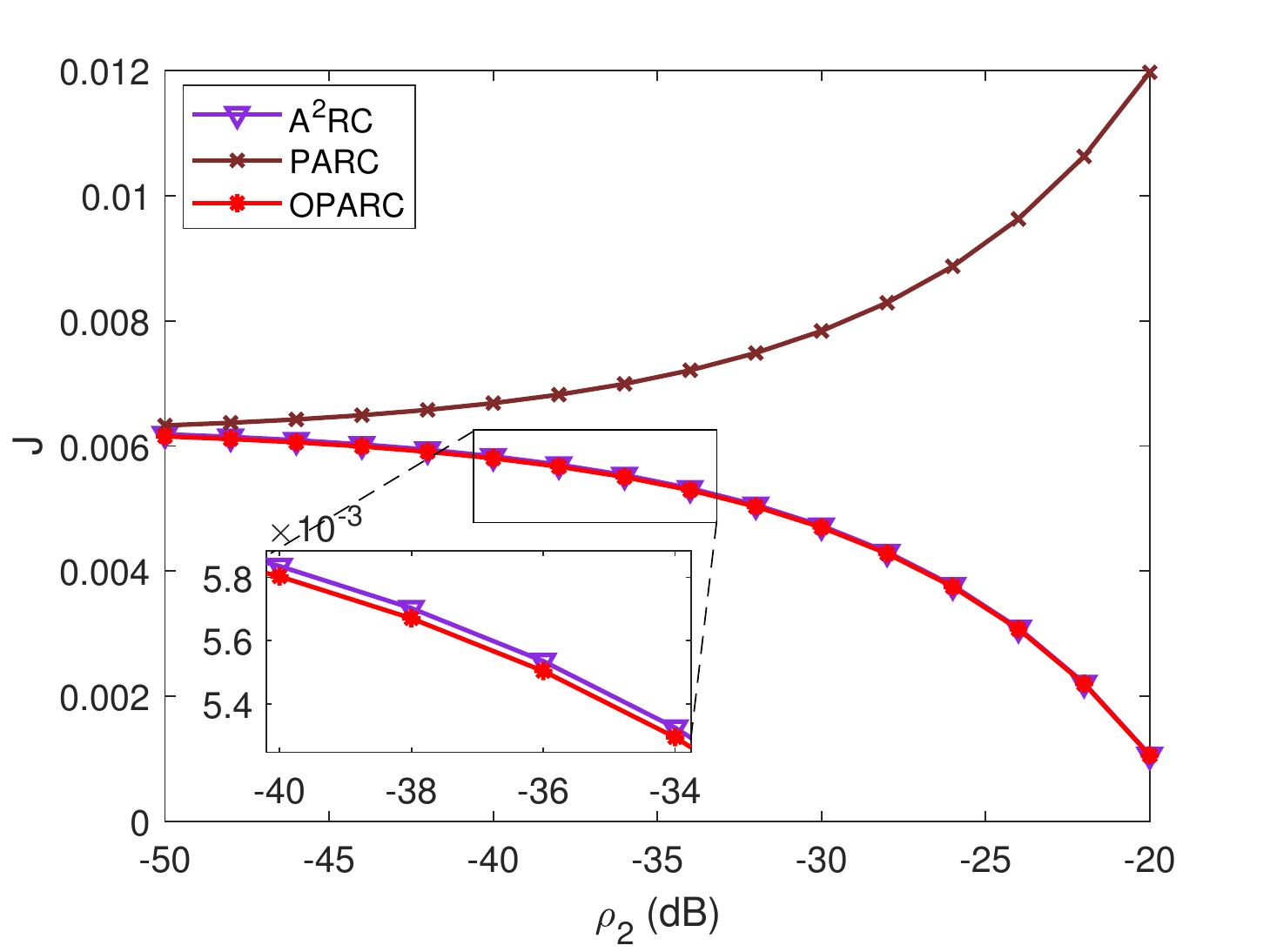}%
		\label{sideJ}}							
	\caption{Pattern variation comparison (the first example).}
	\label{trans00182dqiwang33}
\end{figure*}

\section{Simulation Results}
We next present some simulations to verify the effectiveness of our proposed OPARC.
To validate the superiority of $ \textrm{OPARC} $, we also test 
another precise array response control ($ \textrm{PARC} $) scheme, in which we
adopt the following non-optimal parameter intentionally:
\begin{align}\label{gamatimes}
\gamma_k={\gamma}_{k,\times}
\triangleq\left\{{\gamma_{k,a}},{\gamma}_{k,b}\right\}\setminus{\gamma}_{k,\star}
\end{align}
and use the same remaining procedure as $ \textrm{OPARC} $.
Denote $ {\beta}_{k,\times}=\Psi^{-1}_{k}({\gamma}_{k,\times}) $.
Note that $ {\gamma}_{1,\times} $ is the same as that in Corollary 1.
Clearly, $ \textrm{PARC} $ can precisely control array response level as well, while its parameter $ \gamma_k $ is not optimally selected as in OPARC.
Besides $ \textrm{OPARC} $ and $ \textrm{PARC} $,
the ${\textrm A}^2{\textrm{RC}}$ algorithm in \cite{snrf41} is also compared.
We set $ \omega=6\pi\times10^{8}~{\rm rad/s} $, which corresponds to a wavelength $ \lambda=2\pi c/\omega=1{\rm m} $
with the light speed $ c $.
Consider a 11-element nonuniform spaced linear array
	with nonisotropic elements.
Both the element locations $ x_n $ and the element patterns $ g_n(\theta) $ are listed in Table \ref{table4}, from
which the $ \tau_n(\theta) $ in \eqref{atheta} can be specified as $ \tau_n(\theta)=x_n{\rm sin}(\theta)/c $.
Additionally, we take the quiescent weight vector $ {\bf a}(\theta_0) $ as the initial weight
and fix the beam axis at $ \theta_0=20^{\circ} $ for all experiments conducted.
For convenience, we carry out two steps of the array response control algorithms and
denote the two adjusted angles as $ \theta_1 $ and $ \theta_2 $, respectively.

To measure the performances of different methods, we introduce two cost functions.
The first one is defined as
\begin{align}\label{D}
D\triangleq|L_2(\theta_1,\theta_0)-L_1(\theta_1,\theta_0)|
\end{align}
where $ L_k(\theta,\theta_0) $ represents the resultant response after finishing
the $ k $-th step of weight update.
It is seen that $ D $ measures the response level difference between two consecutive response controls at $ \theta_1 $.
The second cost function is defined as
\begin{align}
J\buildrel \Delta \over =\sqrt{\dfrac{1}{I} \sum\limits^{I}_{i=1}
\begin{vmatrix}L_2(\vartheta_i,\theta_0)-L_1(\vartheta_i,\theta_0)\end{vmatrix}^2}
\end{align}
where $ \vartheta_i $ stands for the $ i $th sampling point in the angle sector,
$ I $ denotes the number of sampling points.
$ J $ measures the deviation between two response patterns $ L_2 $ and $ L_1 $.
We will uniformly sample the region $ [-90^{\circ},90]^{\circ} $ every
$ 0.2^{\circ} $ and hence obtain $ I=901 $ discrete points.
Besides $ D $ and $ J $ above, we also test the obtained array gains of different methods, and
consider pattern variation and pattern distortion for performance comparison.

\subsection{Pattern Variation}
In the first example, we test the performances of different approaches for
sidelobe response control.
More specifically, the normalized responses at $ \theta_1=-45^{\circ} $ and $ \theta_2=-5^{\circ} $ are expected
to be successively adjusted to $ {\rho}_1=-40{\rm dB} $ and $ {\rho}_2=-30{\rm dB} $.

\begin{table}[!t]
	\renewcommand{\arraystretch}{1.3}
	\caption{Element Locations and Element Patterns of the Nonuniform Linear Array}
	\label{table4}
	\centering
	\begin{tabular}{c | c | c || c | c | c}
		\hline
		$ n $&$ x_n $& $ g_n(\theta) $ & $ n $&$ x_n $& $ g_n(\theta) $\\
		\hline
		1&0.00 &$ 1.00{\rm cos}(1.00\theta) $&7 &3.05&$ 1.02{\rm cos}(1.00\theta) $\\
		2&0.45 &$ 0.98{\rm cos}(0.85\theta) $&8 &3.65&$ 1.08{\rm cos}(0.90\theta) $\\
		3&1.00 &$ 1.05{\rm cos}(0.98\theta) $&9 &4.03&$ 0.96{\rm cos}(0.75\theta) $\\
		4 &1.55 &$ 1.10{\rm cos}(0.70\theta) $&10   &4.60&$ 1.09{\rm cos}(0.92\theta) $\\
		5 &2.10&$ 0.90{\rm cos}(0.85\theta) $&11   &5.00&$ 1.02{\rm cos}(0.80\theta) $\\
		6 &2.60&$ 0.93{\rm cos}(0.69\theta) $&   &  & \\
		\hline
	\end{tabular}
\end{table}

In the first step of response control, we can figure out that
$ {\bf c}_{\gamma}=[-0.1704,-0.0315]^{\mT} $, $ d=-8.5231 - j1.5766 $,
$ \gamma_{1,a}=-0.1559 - j0.0288 $ and $ \gamma_{1,b}=-0.1849 - j0.0342 $.
On this basis, we obtain that $ \zeta=1>0 $ and hence choose $ {\gamma}_{1,\star}={\gamma}_{1,a} $
for OPARC and select $ {\gamma}_{1,\times}={\gamma}_{1,b} $ for PARC, according to \eqref{eqn0361}
and \eqref{gamatimes}, respectively.
Additionally, it can be figured out that $ {\bf c}_{\beta}=[-0.1488,0]^{\mT} $ and
	$ R_{\beta}=1.7171 $. We adopt $ {\beta}_{1,\star}={\beta}_{1,r}=1.5683 $ for
	OPARC and take $ {\beta}_{1,\times}={\beta}_{1,l}=-1.8659 $ for PARC.

For ${\textrm A}^2{\textrm{RC}}$, it is found that $ \mu_1={\gamma}_{1,\star}=-0.1559 - j0.0288 $,
which coincides with the result of Corollary 1.
As predicted, one also obtains that $ {\breve{\beta}}_{1,1}=\beta_{1,\star}=1.5683 $.
Fig. \ref{twosidestep1} illustrates the resultant response patterns
of different schemes.
As we can see, all these three approaches are capable of precisely
controlling the array response levels as expected.
Notice also that the result of ${\textrm A}^2{\textrm{RC}}$ is
exactly the same as that of OPARC.

\begin{figure}[!t]
	\centering
	\includegraphics[width=3.55in]{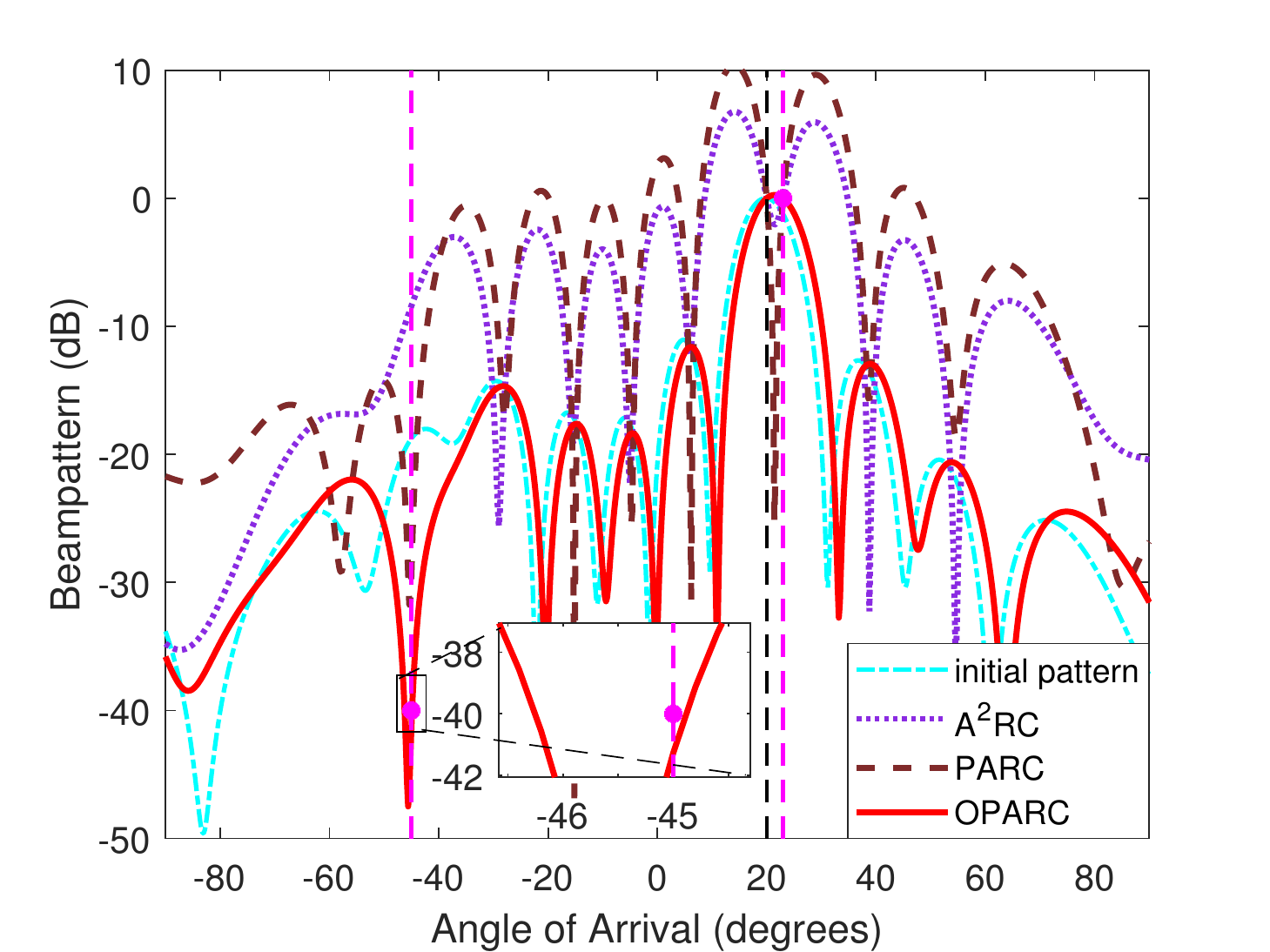}
	\caption{Comparison of synthesized patterns at the 2nd step of the 2nd example.}
	\label{dqiwang4}
\end{figure}


In the second step, with the same manner we found out that
$ {\gamma}_{2,\star}=-0.0685 - j0.0399 $,
$ {\beta}_{2,\star}=0.2504 $,
$ {\gamma}_{2,\times}=-0.1148 - j0.0695 $,
$ {\beta}_{2,\times}=-0.4277 $
and $ \mu_2=-0.0674 - j0.0393 $.
Fig. \ref{twosidestep2} depicts the results of different methods.
It is seen that all methods can adjust $ L(\theta_2,\theta_0) $
to $ \rho_2 $. 
For the proposed OPARC scheme, it can be checked that $ {\beta}_{k,r} $ is
	the ultimate selection of $ {\beta}_{k,\star} $ ($ k=1,2 $). In fact, this is consistent with the
	conclusion of Proposition 5.
	One can see that both $ \beta_{1,\star} $ and $ \beta_{2,\star} $ are positive in this case.
This coincides with the theoretical prediction of Proposition 6, since it is required to lower
the response levels in either step.

To further examine the performance, we evaluate $ D $ and $ J $
as defined earlier and list their measurements in Table \ref{table5}.
It is observed that the proposed OPARC scheme minimizes
both $ D $ and $ J $ among the three methods.
From Table \ref{table5} and Fig. \ref{twosidestep2}, 
it is found
that ${\textrm A}^2{\textrm{RC}}$ causes serious perturbation
(about $ 5{\rm dB} $) at the previous point $ \theta_1 $.
In fact, besides the virtual interference assigned
	at $ \theta_2 $, another one is also assigned at the previously adjusted direction (i.e., $ \theta_1 $),
	for the existing ${\textrm A}^2{\textrm{RC}}$ algorithm.
	We can calculate that $ {\breve{\beta}}_{2,2}=0.2465 + j0.0001 $ and $ {\breve{\Delta}}_{2,1}=-0.4120 + j2.5879$
	(INR of the additional interference assigned at $ \theta_1 $ in the second step
	of response control).
	Notice that both $ {\breve{\beta}}_{2,2} $ and $ {\breve{\Delta}}_{2,1} $ are complex numbers, which is different from that of  $ \textrm{OPARC} $.
Finally, we have listed the obtained array gains of these approaches at both steps in Table \ref{table5}.
Clearly, it is seen that OPARC outperforms the other two methods.

Since both $ D $ and $ J $ depend on the desired level at $ \theta_2 $ (i.e., $ \rho_2 $), we vary $ {\rho}_2 $ from $ -50{\rm dB} $ to $ -20{\rm dB} $ and
recalculate $ D $ and $ J $ with the other settings unchanged.
Fig. \ref{sideD} and Fig. \ref{sideJ} plot the curves of $ D $ versus $ \rho_2 $
and $ J $ versus $ \rho_2 $, respectively.
It can be clearly observed that the proposed OPARC algorithm performs the best
on both $ D $ and $ J $.
The existing ${\textrm A}^2{\textrm{RC}}$ performs well on $ J $, however, it
causes a large deviation to the response level at $ \theta_1 $ as displayed in Fig. \ref{sideD}.

%
\begin{table}[!t]
	\renewcommand{\arraystretch}{1.3}
	\caption{Obtained Parameter Comparison (the First Example)}
	\label{table5}
	\centering
	\begin{tabular}{c | c | c | c }
		\hline
		&${\textrm A}^2{\textrm{RC}}$&${\textrm{PARC}}$&${\textrm{OPARC}}$ \\
		\hline
		$ D ({\rm dB}) $ &$ 5.05 $ & $ 0.86 $ &$ 0.51 $ \\
		\hline		
		$ J $&$ 4.72e^{-3} $ &$ 7.84e^{-3} $  &$ 4.69e^{-3} $ \\
		\hline
		$ G_{1} ({\rm dB}) $&$ 10.0482 $ &$ 10.0331 $  &$ 10.0482 $ \\
		\hline
		$ G_{2} ({\rm dB}) $&$10.0026$   &$ 9.9653 $ &$ 10.0074 $\\
		\hline		
	\end{tabular}
\end{table}

\subsection{Pattern Distortion}
In this part, we shall further show the advantages of the OPARC.
For convenience, we set $ \theta_1 $ and its desired level $ \rho_1 $ the same as the first example,
and then conduct the second step of the response control by taking $ \theta_2=23^{\circ} $ and $ \rho_2=0{\rm dB} $.
Notice that $ \theta_2 $ is in the mainlobe region in this case, and it is required to elevate the response level there.

Clearly, the obtained parameters of the second step are renewed for all methods tested, while the results of
the first step keep unaltered compared to the previous example.
Here, it can be obtained with the ${\textrm A}^2{\textrm{RC}}$ algorithm that $ \mu_2=-0.5931 + j0.8040 $,
$ {\breve{\beta}}_{2,2}=-0.3923 - j0.4011 $ and $ {\breve{\Delta}}_{2,1}=-1.8001 + j0.0334$.
For the PARC algorithm, we obtain $ {\gamma}_{2,\times}=-0.7108 + j0.7171 $ and
$ \beta_{2,\times}=-0.8522 $.
While for OPARC, its parameter satisfies $ \gamma_{2,\star}=\gamma_{2,b}=0.8352 - j0.8438 $
and $ \beta_{2,\star}=\beta_{2,r}=-0.0577$.
It is worth noting that we have selected $ \gamma_b $, which
is different from that at Step 1 (where $ \gamma_a $ is selected),
to obtain the final $ \gamma_{2,\star} $.
In fact, this flexible mechanism of parameter determination in OPARC
enables us to avoid certain pattern distortion, which is inevitable
in ${\textrm A}^2{\textrm{RC}}$ or PARC.
To see this clearer, we have depicted the synthesized patterns in Fig. \ref{dqiwang4}.
It can be found that all the response levels at $ \theta_2 $
still meet the requirement as before.
However, it shows clearly that the patterns of ${\textrm A}^2{\textrm{RC}}$ and PARC are
severely distorted. 
The obtained mainlobes are split
and the resultant sidelobe levels are raised for both ${\textrm A}^2{\textrm{RC}}$ and PARC.
For the proposed OPARC, none of the above
undesirable phenomena happens and a well-shaped pattern has been obtained.
By the way, notice that $ \beta_{2,\star} $ is negative in this scenario.
This is consistent with the conclusion of Proposition 6, since the response level needs to be lifted in
this second step of response control.

\begin{figure}[!tpb]
	\centering\subfloat[Curves of $ D $ versus $ {\rho}_2 $]
	{\includegraphics[width=3.55in]{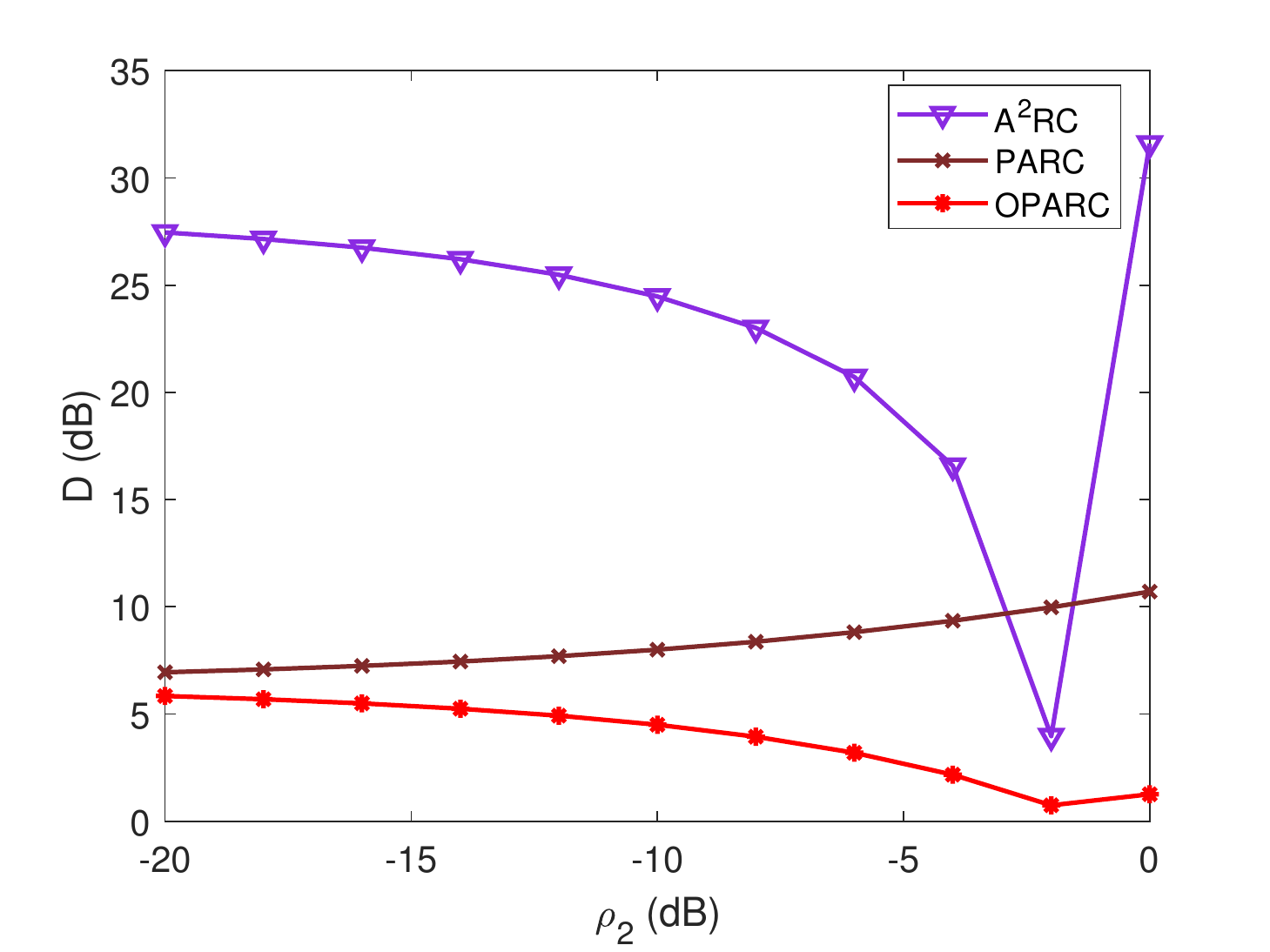}%
		\label{sideandmainD}}\
	\centering\subfloat[Curves of $ J $ versus $ {\rho}_2 $]
	{\includegraphics[width=3.55in]{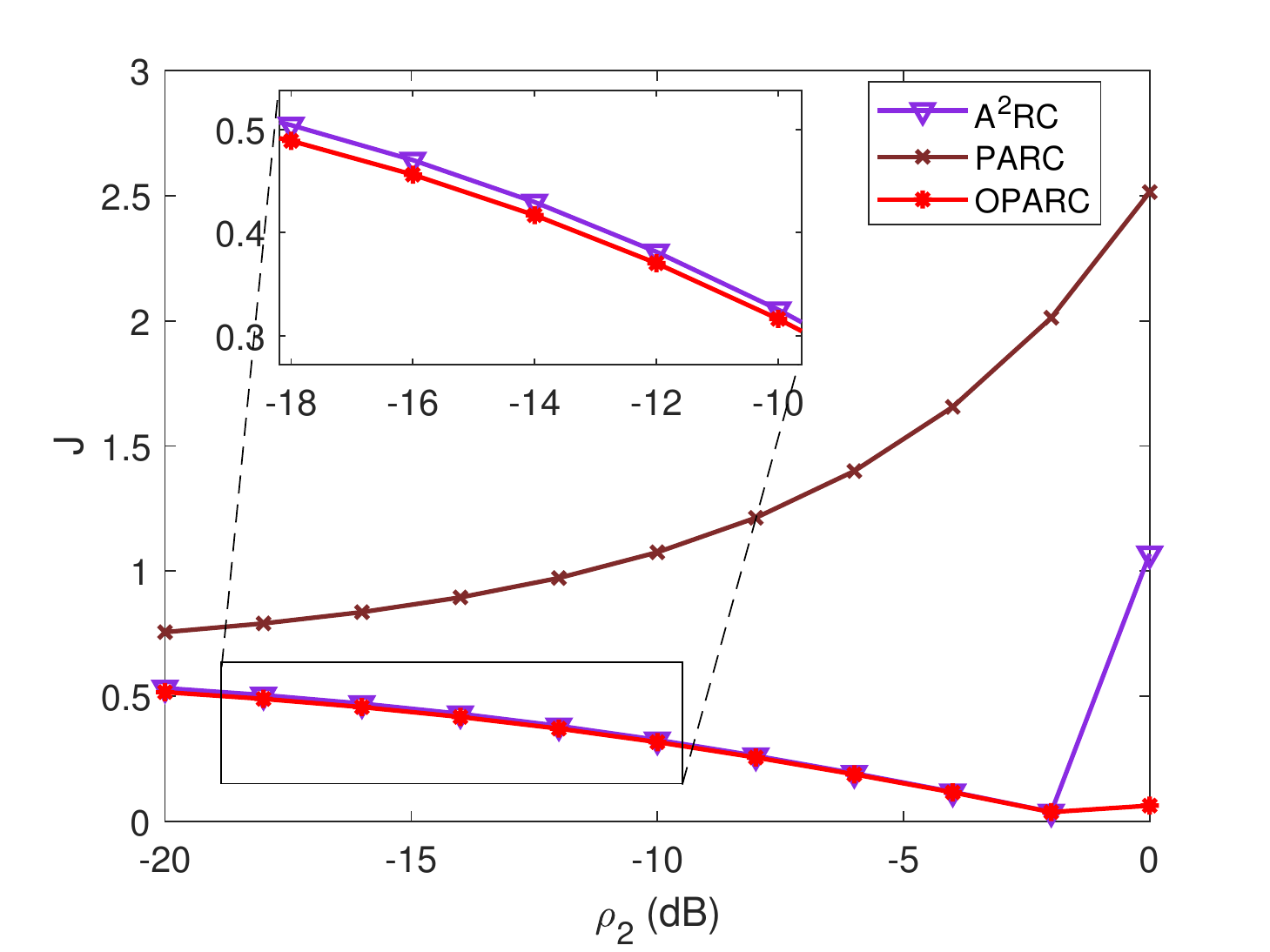}%
		\label{sideandmainJ}}			
	\caption{Pattern variation comparison (the second example).}
	\label{dqiwang333}
\end{figure}

The details of $ D $, $ J $ and the obtained array gains have been specified in Table \ref{table65},
from which the merits of the proposed OPARC algorithm are clearly observed.
Note that the array gain $ G_1 $ is not listed in Table \ref{table65} since it has been reported in Table \ref{table5}.
Again, to further examine the performance, we vary $ {\rho}_2 $ from $ -20{\rm dB} $
to $ 0{\rm dB} $, and depict the curve of $ D $ versus $ {\rho}_2 $ in Fig. \ref{sideandmainD} and
curve of $ J $ versus $ \rho_2 $ in Fig. \ref{sideandmainJ}, respectively.
As illustrated in these two figures, ${\textrm A}^2{\textrm{RC}}$ causes a great
perturbation on $ \theta_1 $  (i.e., high value of $ D $) and PARC performs poor on the average deviation $ J $.
On the other hand, OPARC performs the best when measuring either $ D $ or $ J $.

\begin{table}[!t]
	\renewcommand{\arraystretch}{1.3}
	\caption{Obtained Parameter Comparison (the Second Example)}
	\label{table65}
	\centering
	\begin{tabular}{c | c | c | c }
		\hline
		&${\textrm A}^2{\textrm{RC}}$&${\textrm{PARC}}$&${\textrm{OPARC}}$ \\
		\hline
		$ D ({\rm dB}) $ &$ 31.6001 $ &$ 10.7083 $ &$ 1.2595 $ \\
		\hline		
		$ J $&$ 1.0685  $ &$ 2.5149 $  &$ 0.0624 $ \\
		\hline
		$ G_{2} ({\rm dB}) $&$2.5060$   &$ 0.7366 $ &$ 13.1370 $\\
		\hline		
	\end{tabular}
\end{table}

\begin{figure*}[!t]
	\centering
	\subfloat[$ {{\Re}(d)}>{{\bf c}_{\gamma}(1)}\geq0$]
	{\includegraphics[width=1.75in]{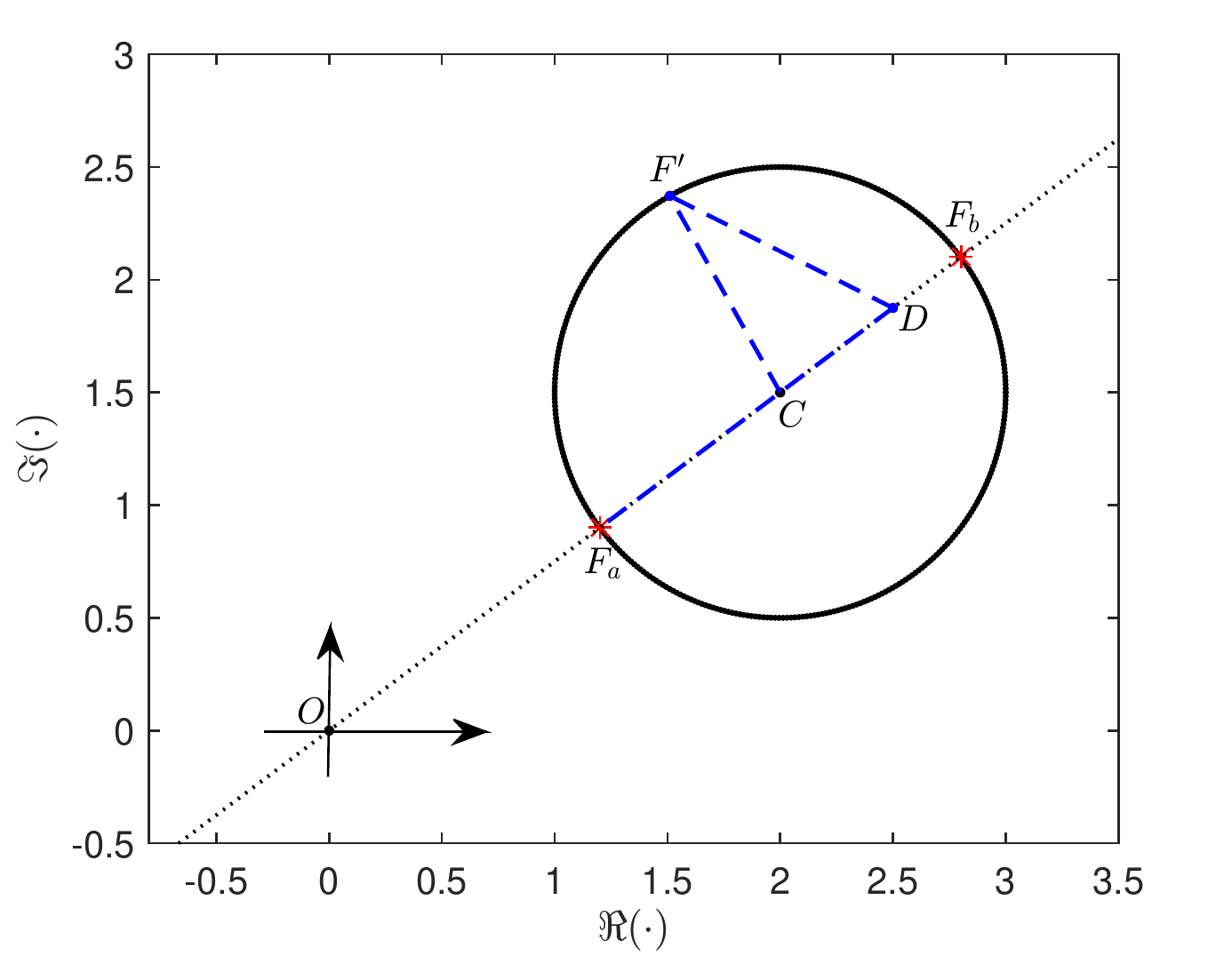}%
		\label{Modifulupic2}}
	\hfil
	\subfloat[$ {{\Re}(d)}<{{\bf c}_{\gamma}(1)}<0$]
	{\includegraphics[width=1.75in]{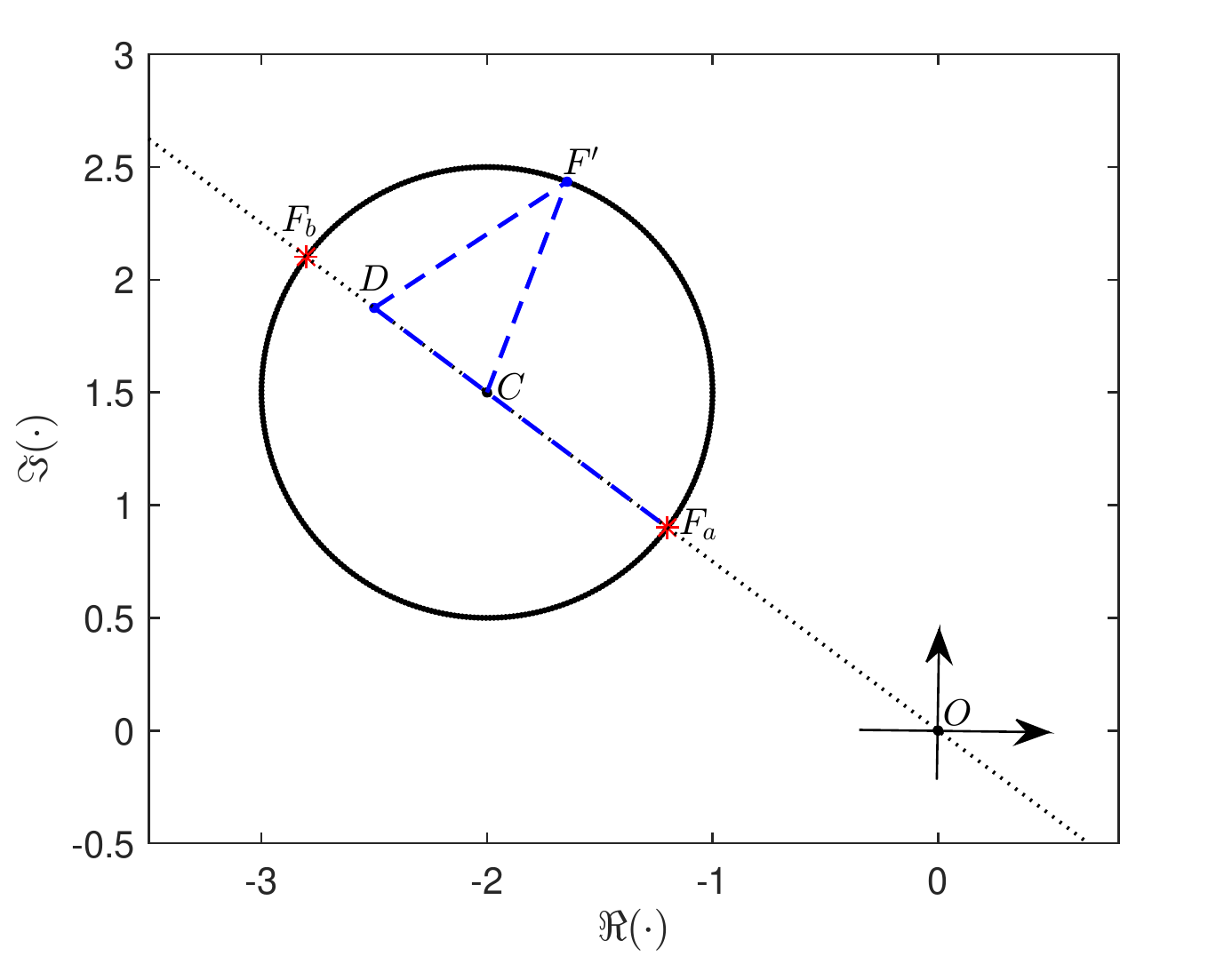}%
		\label{Modifulupic4}}
	\hfil
	\subfloat[$ {\bf c}_{\gamma}(1)\geq0 $ and $ {{\Re}(d)}<{{\bf c}_{\gamma}(1)}$]
	{\includegraphics[width=1.75in]{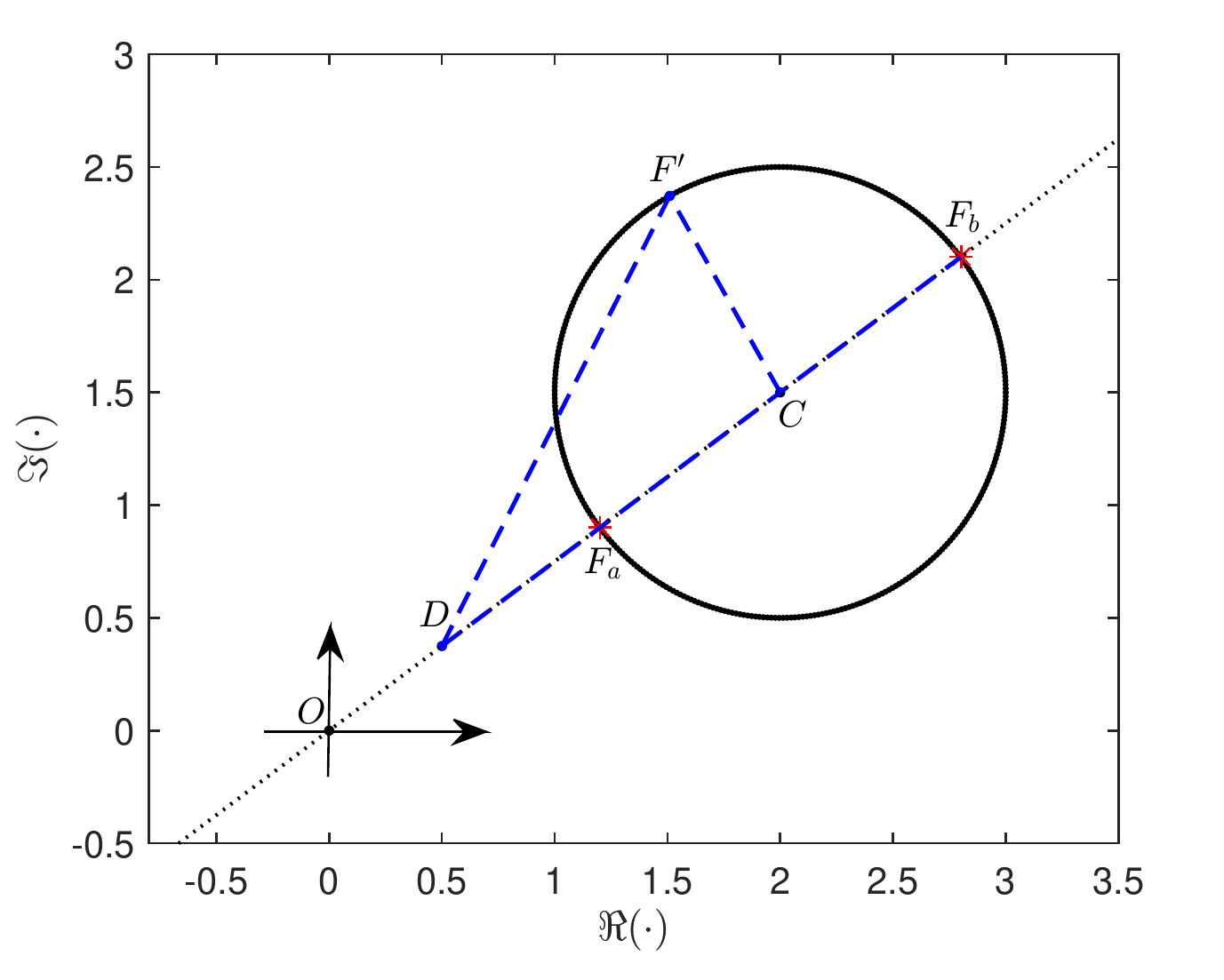}%
		\label{Modifulupic1}}
	\hfil
	\subfloat[$ {\bf c}_{\gamma}(1)<0 $ and $ {{\Re}(d)}>{{\bf c}_{\gamma}(1)}$]
	{\includegraphics[width=1.75in]{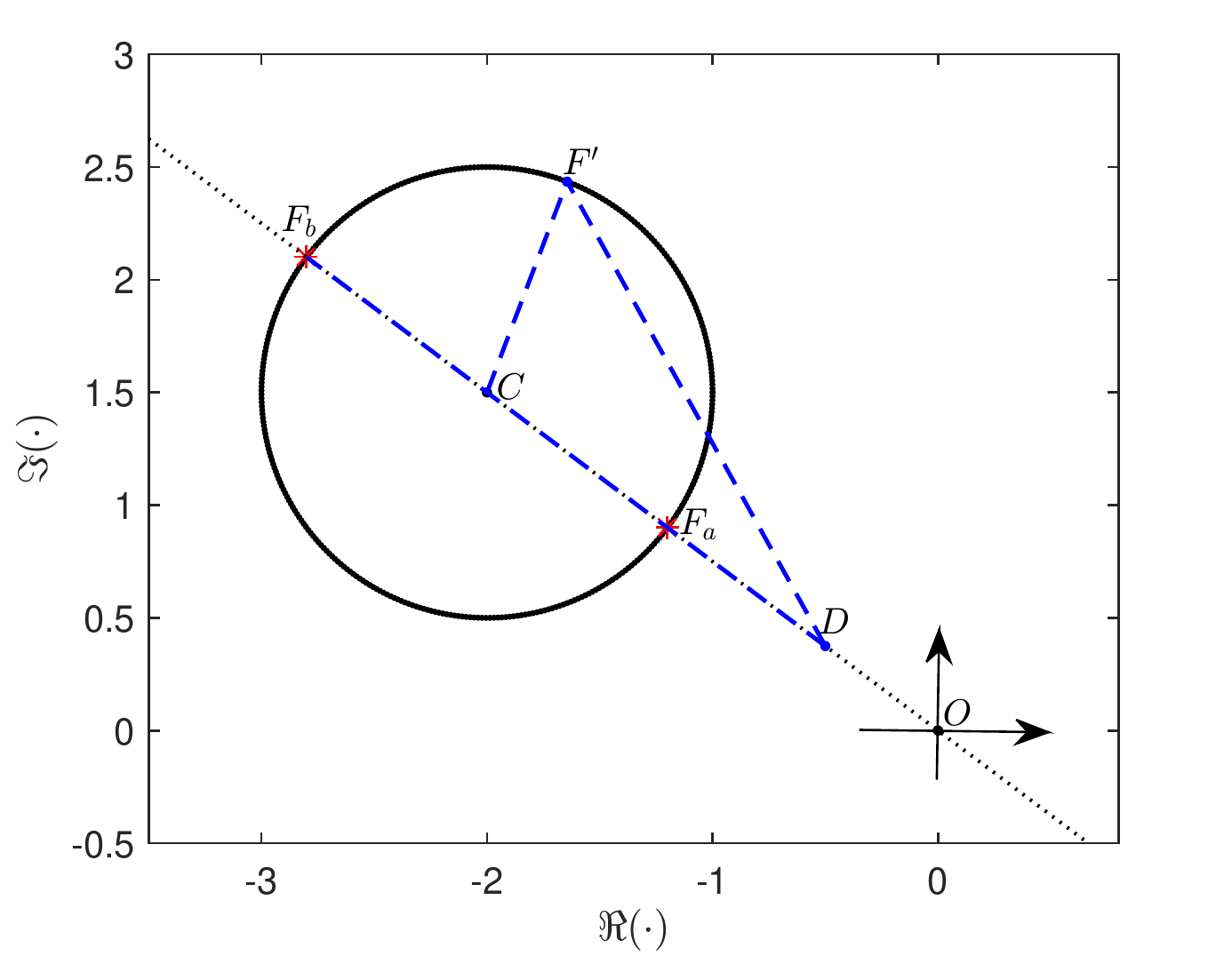}%
		\label{Modifulupic3}}			
	\caption{Geometrical illustration of different cases when maximizing array gain.}
	\label{Modifulupic}
\end{figure*}

\section{Conclusions}
In this paper, a novel algorithm of optimal and precise array response control (OPARC) has been proposed.
This algorithm origins from the adaptive array theory
and the change rule of the optimal weight vector, when adding interferences one by one, has been found.
Then, the parameter selection mechanism has been carried out to
maximize the array gain with the constraint that the response level at one
direction is precisely adjusted.
Some properties of OPARC have been presented and OPARC is compared in details with $ {\textrm A}^2\textrm{RC} $.
Finally, simulation results have been shown to illustrate the effectiveness of the proposed OPARC method.
Based on the fundamentals developed in this paper,
a further
extension of OPARC
to multi-point array response control and its applications to, for example,
pattern synthesis, multi-constraint adaptive beamforming and
quiescent pattern control will be considered in \cite{p3}.

\appendices

\section{Proof of Proposition 2}
Since $ {\bf T}_{k-1} $ is
assumed to be Hermitian, both $ {\xi}_0 $ and $ {\xi}_k $ are real-valued
and one also gets $ \widetilde{{\xi}}_c={\xi}^{*}_c $. According to \eqref{eqn0013}, we have
\begin{align}
{\bf H}_k(1,2)={\bf w}^{\mH}_{k-1}
[{\bf a}(\theta_k){\bf a}^{\mH}(\theta_k)-{\rho}_k{\bf a}(\theta_0){\bf a}^{\mH}(\theta_0)
]{\bf v}_k=\chi{\xi}^{*}_c\nonumber
\end{align}
where
$ \chi={\xi}_k-{\rho_k}{\xi}_0
\in \mathbb{R}$.
From \eqref{eqn0030}, we have
\begin{align}\label{eqn366}
{\bf c}_{\gamma}(1)+j{\bf c}_{\gamma}(2)
=-{\chi{\xi}_c}/{{\bf H}_k(2,2)}.
\end{align}
Thus,
\begin{align}
{\gamma}_{k,a}
=-\dfrac{\left(\|{\bf c}_{\gamma}\|_2-R_{\gamma}\right)\chi{\xi}_c}
{\|{\bf c}_{\gamma}\|_2{\bf H}_k(2,2)},~
{\gamma}_{k,b}
=-\dfrac{\left(\|{\bf c}_{\gamma}\|_2+R_{\gamma}\right)\chi{\xi}_c}
{\|{\bf c}_{\gamma}\|_2{\bf H}_k(2,2)}.\nonumber
\end{align}
From \eqref{eqn366} and both $ \chi $ and $ {{\bf H}_k(2,2)} $ are real, we have
\begin{align}\label{f019}
{{\bf c}_{\gamma}(2)}\big/{{\bf c}_{\gamma}(1)}
=
{{\Im}({\xi}_c)}\big/
{{\Re}({\xi}_c)}.
\end{align}

For the array gain $ G_k $ in \eqref{defG} we have
\begin{align}\label{f020}
G_k&=|{\xi}^{*}_c|\cdot
|{{\xi}_0}/{\xi}^{*}_c+{\gamma}_k|=|{\xi}^{*}_c|\cdot
\big|{\gamma}_k-d
\big|\nonumber\\
&=|{\xi}^{*}_c|\cdot
\big\|\begin{bmatrix}
{\Re}(\gamma_k)&{\Im}(\gamma_k)
\end{bmatrix}^{\mT}-
\begin{bmatrix}
{\Re}(d)&{\Im}(d)
\end{bmatrix}^{\mT}
\big\|_2
\end{align}
where $ d=-{{\xi}_0}/{{\xi}^{*}_c} $.
Also,
\begin{align}\label{f024}
{{\Im}(d)}\big/{{\Re}(d)}
={{\Im}({\xi}_c)}\big/
{{\Re}({\xi}_c)}
={{\bf c}_{\gamma}(2)}\big/{{\bf c}_{\gamma}(1)}.
\end{align}
This shows that the origin $ {\bf O} $, the center $ {\bf c}_{\gamma} $
and $ D=[{\Re(d)}~{\Im(d)}] $ are co-linear on the plane as shown in Fig. \ref{Modifulupic}.
Note that in Fig. \ref{Modifulupic} we have denoted $ C $ as the center of the circle.
From \eqref{f020} it can be observed that $ G_k $ is a scaling
of the Euclidean distance between $ D $ and
a point $ F=[\Re(\gamma_k),\Im(\gamma_k)]^{\mT} $
located on the circle $ \mathbb{C}_{\gamma} $.
From this observation, the optimal solution to \eqref{qu011} can thus be
obtained in a geometrical approach below.

Without loss of generality, we first assume that $ {{\Re}(d)}\neq{{\bf c}_{\gamma}(1)} $,
otherwise, all $ {\gamma_k} $ on circle $ \mathbb{C}_k $ will have the
same $ G_k $. In the case of $ {{\Re}(d)}>{{\bf c}_{\gamma}(1)}\geq0$ or $ {{\Re}(d)}<{{\bf c}_{\gamma}(1)}<0$, it can be
derived that $ {\gamma}_{k,\star}={\gamma}_{k,a} $, i.e.,
when $ F=F_a $, $ G_k $	is maximized.
In fact, these two cases can be geometrically illustrated by Fig. \ref{Modifulupic2}
and Fig. \ref{Modifulupic4}, respectively.

Similarly, in the case of $ {{\bf c}_{\gamma}(1)}\geq0 $,
$ {{\Re}(d)}<{{\bf c}_{\gamma}(1)} $ (as shown in Fig. \ref{Modifulupic1}), or $ {{\bf c}_{\gamma}(1)}<0 $,
$ {{\Re}(d)}>{{\bf c}_{\gamma}(1)} $
(as shown in Fig. \ref{Modifulupic3}), the two points $ {O} $ and $ D $ are located on the same sides (right or left) of $ C $. As a result, $ {\gamma}_{k,\star}={\gamma}_{k,b} $,
i.e., when $ F=F_b $, $ G_k $ is maximized in these two cases.

In summary, it can be concluded that
if $ \zeta>0 $, $ {\gamma}_{k,\star}={\gamma}_{k,a} $,
otherwise, $ {\gamma}_{k,\star}={\gamma}_{k,b} $,
where $ \zeta $ has been defined in \eqref{zetadef}.
This completes the proof.

\section{Proof of Proposition 5}
It is easy to see that
$ {-1}/{\xi_k}>{{\xi}_0}/\left({|{\xi}_c|^2-{{\xi}_0}{{\xi}_k}}\right) $ in \eqref{qua011} is
actually equivalent to
\begin{align}\label{eqn2063}
({|{\xi}_c|^2-{{\xi}_0}{{\xi}_k}}){{\xi}_k}<0.
\end{align}

When $ {\bf T}_{k-1}\!\in\!\mathbb{S}^N_{++} $,
we have $ {\bf T}^{-1}_{k-1}\!\!\in\!\mathbb{S}^N_{++} $.
Thus, from \eqref{20b}
\begin{align}\label{eqn2065}
{{\xi}_k}>0.
\end{align}
Let the Cholesky decomposition of $ {{\bf T}^{-1}_{k-1}} $ be
$ {{\bf T}^{-1}_{k-1}}={\bf \Xi}{\bf \Xi}^{\mH} $,
where $ {\bf \Xi} $ is an invertible matrix.
If $ {\bf a}(\theta_0)\neq\varrho{\bf a}(\theta_k) $ for $ \forall{\varrho}\in{\mathbb{C}} $ that always holds in array antenna theory, we have
$ {\bf \Xi}^{\mH}{\bf a}(\theta_0)\neq\varrho{\bf \Xi}^{\mH}{\bf a}(\theta_k) $ for $ \forall{\varrho}\in{\mathbb{C}} $.
Then, from the Cauchy-Schwarz inequity we have
\begin{align}\label{eqn2067}
{|{\xi}_c|^2-{{\xi}_0}{{\xi}_k}}&=
\big|({\bf \Xi}^{\mH}{\bf a}(\theta_k))^{\mH}({\bf \Xi}^{\mH}{\bf a}(\theta_0))
\big|^2-\nonumber\\
&~~~~~~~~~\big\|{\bf \Xi}^{\mH}{\bf a}(\theta_0)\big\|^2_2\cdot\big\|{\bf \Xi}^{\mH}{\bf a}(\theta_k)\big\|^2_2<0.
\end{align}
Hence, from Proposition 4, we have $ {\beta}_{k,\star}={\beta}_{k,r} $.

From \eqref{eqn053}, \eqref{qua015} and the fact that $ {\beta}_{k,\star}={\beta}_{k,r} $, we have
\begin{align}
{\beta}_{k,\star}=\dfrac{|\xi_c|-\sqrt{{\rho}_k}{\xi}_0}{\sqrt{{\rho}_k}({{\xi}_0}{{\xi}_k}-|{\xi}_c|^2)}.
\end{align}
This completes the proof of \eqref{qua017}.

Furthermore, if $ {\bf T}_{k-1}\in\mathbb{S}^N_{++} $, 
one learns from $ {\bf T}_k={\bf T}_{k-1}+{\beta}_{k,\star}{\bf a}(\theta_k){\bf a}^{\mH}(\theta_k) $ that
\begin{subequations}	
	\begin{align}
\!{\bf T}_k\in\mathbb{S}^N_{++}&\Leftrightarrow{\bf T}_{k-1}+{\beta}_{k,\star}{\bf a}(\theta_k){\bf a}^{\mH}(\theta_k)\in\mathbb{S}^N_{++}\\
	&\Leftrightarrow{\bf I}\!+\!{\beta}_{k,\star}{\bf T}^{-1/2}_{k-1}{\bf a}(\theta_k){\bf a}^{\mH}(\theta_k)\!{\bf T}^{-1/2}_{k-1}\!\in\!\mathbb{S}^N_{++}\\
	&\Leftrightarrow 1+{\beta}_{k,\star}{\bf a}^{\mH}(\theta_k){\bf T}^{-1}_{k-1}{\bf a}(\theta_k)>0\\
\label{di0050}&\Leftrightarrow{\beta}_{k,\star}>-1/{\xi}_k\\
	&\Leftrightarrow\dfrac{|{\xi}_c|{\xi_k}-\sqrt{\rho_k}|{\xi}_c|^2}{{\sqrt{{\rho}_k}({{\xi}_0}{{\xi}_k}-|{\xi}_c|^2)}{\xi}_k}>0\\
	&\Leftrightarrow{\rho}_k<{\xi}^2_k/|{\xi}_c|^2.
	\end{align}
\end{subequations}
This completes the proof of Proposition 5.

\section{Proof of Proposition 7}
From the update procedure of weight vector, one gets
\begin{align}\label{modi0050}
{\bf w}_k&={\bf w}_{k-1}+{\gamma}_k{\bf v}_{k}\nonumber\\
&={\bf T}^{-1}_{k-1}{\bf a}(\theta_0)-{\beta_k}{\xi_c}{\bf T}^{-1}_{k-1}{\bf a}(\theta_k)/(1+{\beta_k}\xi_k).
\end{align}
Then we can obtain $ {\bf a}^{\mH}(\theta_0){\bf w}_k={\xi_0}-{\beta_k}|{\xi_c}|^2/(1+{\beta_k}\xi_k) $ and
\begin{align}\label{modi0151}
|{\bf w}^{\mH}_k{\bf a}(\theta_0)|^2=|
(\xi_0\xi_k-|\xi_c|^2){\beta_k}+{\xi_0}
|^2\big/|1+{\beta}_k{\xi_k}|^2.
\end{align}
On the other hand, from \eqref{modi0050} we have
$ {\bf w}^{\mH}_k{\bf T}_{k-1}={\bf a}^{\mH}(\theta_0)-{\beta}^{\ast}_{k}{\xi}^{\ast}_c{\bf a}^{\mH}(\theta_k)/
(1+{\beta}^{\ast}_{k}{\xi_k}) $ and further get
\begin{align}\label{modi0152}
{{\bf w}^{\mH}_{k}{\bf T}_{k-1}{\bf w}_{k}}&=
\left(
{\bf a}^{\mH}(\theta_0)-{\beta}^{\ast}_{k}{\xi}^{\ast}_c{\bf a}^{\mH}(\theta_k)/
(1+{\beta}^{\ast}_{k}{\xi_k})
\right)\cdot\nonumber\\
&~~\left(
{\bf T}^{-1}_{k-1}{\bf a}(\theta_0)-{\beta_k}{\xi_c}{\bf T}^{-1}_{k-1}{\bf a}(\theta_k)/(1+{\beta_k}\xi_k)
\right)\nonumber\\
&=\dfrac{(\xi_0\xi_k-|\xi_c|^2){\xi_k}|{\beta_k}+\frac{1}{{\xi_k}}|^2+
	\frac{|\xi_c|^2}{{\xi_k}}}{|1+{\beta}_k{\xi_k}|^2}.
\end{align}
Combining \eqref{modi0151} and \eqref{modi0152}, we get
\begin{align}\label{modi0155}
\dfrac{|{\bf w}^{\mH}_k{\bf a}(\theta_0)|^2}
{{\bf w}^{\mH}_{k}{\bf T}_{k-1}{\bf w}_{k}}=
\dfrac{\left(\frac{\xi_0\xi_k-|\xi_c|^2}{\xi_k}\right)\cdot{R^2_{\beta}}}
{\big|{\beta}_k+\frac{1}{{\xi_k}}\big|^2+\frac{|\xi_c|^2}{(\xi_0\xi_k-|\xi_c|^2){\xi^2_k}}}
\end{align}
where we have utilized the fact that $ \xi_0\xi_k-|\xi_c|^2>0 $ (see \eqref{eqn2067}
when $ {\bf T}^{-1}_{k-1}\in\mathbb{S}^N_{++} $) and
$ \big|{\beta}_k+\frac{\xi_0}{\xi_0\xi_k-|\xi_c|^2}\big|=R_{\beta} $.
Obviously, the maximization of $ {|{\bf w}^{\mH}_k{\bf a}(\theta_0)|^2}/
({{\bf w}^{\mH}_{k}{\bf T}_{k-1}{\bf w}_{k}}) $ is equivalent to minimizing
$ \big|{\beta}_k+\frac{1}{{\xi_k}}\big| $. 
Define
\begin{align}\label{qu561}
{\bf f}\triangleq\begin{bmatrix}-1/{{\xi}_k}&0\end{bmatrix}^{\mT}
\end{align}
then we can reformulate problem \eqref{qu552}
as
\begin{subequations}\label{qu511}	
	\begin{align}
	\mathop {{\rm{minimize}}}\limits_{{\beta _k}}&~~~{\|[{\Re}(\beta_k)~~{\Im}(\beta_k)]^{\mT}-{\bf f}\|_2}\\
	\label{qu611}{\rm subject~to}&~~~[{\Re}(\beta_k)~~{\Im}(\beta_k)]^{\mT}\in{\mathbb{C}_{\beta}}.
	\end{align}
\end{subequations}

\newcounter{MYtempeqncnt}
\begin{figure*}[t]
	\normalsize
	\setcounter{MYtempeqncnt}{\value{equation}}
	\setcounter{equation}{80}
	\begin{align}\label{xip}
	{\bf T}^{-1}_{p}{\bf a}(\theta_r)={\bf T}^{-1}_{p-1}{\bf a}(\theta_r)+\nu{\bf T}^{-1}_{p-1}{\bf a}(\theta_p)
	&={\bf A}(\theta_r,\theta_{p-1},\cdots,\theta_{1}){\bf d}_p(\theta_r)+
	\nu{\bf A}(\theta_p,\theta_{p-1},\cdots,\theta_{1}){\bf d}_p(\theta_p)\nonumber\\
	&={\bf A}(\theta_{r},\theta_{p},\cdots,\theta_{1}){\bf d}_{p+1}(\theta_r),~{\forall}\theta_r\in\mathbb{R}
	\end{align}
	\setcounter{equation}{\value{MYtempeqncnt}}
	\hrulefill
	\vspace*{1pt}
\end{figure*}
\begin{figure*}[t]
	\normalsize
	\setcounter{MYtempeqncnt}{\value{equation}}
	\setcounter{equation}{82}
	\begin{align}\label{dr}
	{\bf d}_{p+1}(\theta_r)=\left[
	{\bf d}_{p}(\theta_r)_1,0,{\bf d}_{p}(\theta_r)_2,\cdots,{\bf d}_{p}(\theta_r)_p
	\right]^{\mT}+
	\nu\left[
	0,{\bf d}_{p}(\theta_p)_p,{\bf d}_{p}(\theta_r)_{p-1},\cdots,{\bf d}_{p}(\theta_r)_1\right]^{\mT}
	\end{align}
	\setcounter{equation}{\value{MYtempeqncnt}}
	\hrulefill
	\vspace*{1pt}
\end{figure*}

\begin{figure*}[!t]
	\normalsize
	\setcounter{MYtempeqncnt}{\value{equation}}
	\setcounter{equation}{84}		
	\begin{align}\label{eqn158}
	{\bf 1}\oslash{\rm diag}({\breve{\bf \Sigma}}_k)=-({\bf A}^{\mH}_k{\breve{\bf w}}_{k})\oslash{\bf b}_{k}
	=-\begin{bmatrix}
	{\bf A}^{\mH}_{k-1}{\breve{\bf w}}_{k}\\{\bf a}^{\mH}(\theta_k){\breve{\bf w}}_{k}
	\end{bmatrix}\oslash
	\begin{bmatrix}
	{\bf b}_{k-1}\\{\mu}_k
	\end{bmatrix}
	&=-\begin{bmatrix}
	({\bf A}^{\mH}_{k-1}{\breve{\bf w}}_{k-1})\oslash{\bf b}_{k-1}+
	({\mu}_k{\bf A}^{\mH}_{k-1}{\bf a}(\theta_k))\oslash{\bf b}_{k-1}\\
	{\bf a}^{\mH}(\theta_k){\breve{\bf w}}_{k-1}/{\mu}_k+\|{\bf a}(\theta_k)\|^2_2\end{bmatrix}\nonumber\\
	&=\begin{bmatrix}
	{\bf 1}\oslash{\rm diag}({\breve{\bf \Sigma}}_{k-1})-
	({\mu}_k{\bf A}^{\mH}_{k-1}{\bf a}(\theta_k))\oslash{\bf b}_{k-1}\\
	-{\bf a}^{\mH}(\theta_k){\breve{\bf w}}_{k-1}/{\mu}_k-\|{\bf a}(\theta_k)\|^2_2\end{bmatrix}.
	\end{align}
	\setcounter{equation}{\value{MYtempeqncnt}}
	\hrulefill
	\vspace*{1pt}
\end{figure*}

On the other hand, substituting the constraint \eqref{qu212} into $ G_k $ and recalling the conclusion of Proposition 3, the array gain
satisfies
\begin{subequations}\label{eqn69}
	\begin{align}
	\label{eqn80}G_k&=\big|{\bf a}^{\mH}(\theta_0){\bf T}^{-1}_{k}{\bf a}(\theta_0)\big|\\
	\label{eqn70}&=\Big|{\xi_0}-\dfrac{\beta_k|{\xi_c}|^2}{1+{\beta_k}{\xi_k}}\Big|\\
	\label{eqn72}&=\Big|\dfrac{{\xi_0}{\xi_k}-|{\xi_c}|^2}{{\xi_k}}\Big|\cdot
	\Big|
	\dfrac{{\beta_k}-{{\xi}_0}/\left({|{\xi}_c|^2-{{\xi}_0}{{\xi}_k}}\right)}{{\beta_k}-\left(-1/{{\xi}_k}\right)}
	\Big|\\
	\label{eqn73}&=\Big|\dfrac{{\xi_0}{\xi_k}-|{\xi_c}|^2}{{\xi_k}}\Big|\cdot
	\dfrac{\|[{\Re}(\beta_k)~~{\Im}(\beta_k)]^{\mT}-{\bf c}_{\beta}\|_2}
	{\|[{\Re}(\beta_k)~~{\Im}(\beta_k)]^{\mT}-{\bf f}\|_2}\\
	\label{eqn77}&=\dfrac{\big|\left({\xi_0}{\xi_k}-|{\xi_c}|^2\right)\cdot{R_{\beta}}/{\xi_k}\big|}
	{\|[{\Re}(\beta_k)~~{\Im}(\beta_k)]^{\mT}-{\bf f}\|_2}.
	\end{align}
\end{subequations}
Note that \eqref{eqn80} comes from the intermediate result of \eqref{defG}, whereas \eqref{eqn73} is obtained from the result of Proposition 3.
Since $ {\big|\left({\xi_0}{\xi_k}-|{\xi_c}|^2\right)\cdot{R_{\beta}}/{\xi_k}\big|} $ is a constant, 
from \eqref{eqn77}, we can
also reformulate problem \eqref{p2qu011} as \eqref{qu511}. 

Consequently, if $ {\bf T}_{k-1}\in\mathbb{S}^N_{++} $, problem \eqref{p2qu011} has the same optimal solution as 
the problem \eqref{qu552}. This completes the proof.

\section{Derivation of \eqref{qua1}}
We first show 
\begin{align}\label{qua3}
{\bf T}^{-1}_{k-1}{\bf a}(\theta_t)={{\bf A}(\theta_t,\theta_{k-1},\cdots,\theta_{1}){\bf d}_k}(\theta_t),~
{\forall}\theta_t\in\mathbb{R}
\end{align}
where the first component of $ {\bf d}_k(\theta_t) $ is 1.

We use induction to prove \eqref{qua3}. 
When $ k=1 $, since $ {\bf T}_0={\bf I} $,
\eqref{qua3} is obvious, where $ {\bf d}_1(\theta_t) $ degenerates
to the scalar $ 1 $.

Suppose \eqref{qua3} is true when $ k=p $, i.e.,
\begin{align}\label{step1}
{\bf T}^{-1}_{p-1}{\bf a}(\theta_s)={\bf A}(\theta_s,\theta_{p-1},\cdots,\theta_{1}){\bf d}_p(\theta_s),~
{\forall}\theta_s\in\mathbb{R}
\end{align}
where $ {\bf d}_p(\theta_s) $ is a $ p\times1 $ vector with its first entry 1. 

When $ k=p+1 $, we want to show
\begin{align}\label{step2}
{\bf T}^{-1}_{p}{\bf a}(\theta_{r})={\bf A}(\theta_{r},\theta_{p},\cdots,\theta_{1}){\bf d}_{p+1}
(\theta_r),~{\forall}\theta_r\in\mathbb{R}
\end{align}
where ${\bf d}_{p+1}(\theta_r) $ is a $ (p+1)\times1 $ vector with its first entry 1. 

To see \eqref{step2}, one recalls \eqref{modi0008} with $ k=p $, and \eqref{step1}, and obtains \eqref{xip} on the top of
next page, where
\addtocounter{equation}{1}
\begin{align}
\nu=-
{\beta_p{\bf a}^{\mH}(\theta_p){\bf T}^{-1}_{p-1}{\bf a}(\theta_r)}/
[{1+{\beta_p}{\bf a}^{\mH}(\theta_p){\bf T}^{-1}_{p-1}{\bf a}(\theta_p)}]
\end{align}
$ {\bf d}_{p+1}(\theta_r) $ is a $ (p+1)\times 1 $ vector as shown in \eqref{dr},
where $ {\bf d}_{p}(\theta_r)_i $ stands for the $ i $th element of $ {\bf d}_{p}(\theta_r) $.

Then, \eqref{qua1} can be seen by substituting \eqref{qua3} with $\theta_t=\theta_k$ into \eqref{qu5212}.

\section{Proof of Corollary 1}
In the first step of the weight vector update in the OPARC, we have
$ {\mathbb{C}}_{\gamma}={\mathbb{C}}_{\mu} $ due to the fact that $ {\bf T}_0={\bf I} $ and
hence $ {\bf H}_1={\bf Q}_1 $. Therefore, one gets
$ {\mu}_{1,\star}={\gamma}_{1,a} $
since $ [{\Re}(\gamma_{1,a})~ {\Im}(\gamma_{1,a})]^{\mT} $ has the minimum module among 
the elements in the set $ {\mathbb{C}}_{\gamma} $ as shown in Fig. \ref{controlall}.

On the other hand, substituting $ {\bf v}_1={\bf a}(\theta_1) $ and $ {\bf w}_0={\bf a}(\theta_0) $ into
\eqref{eqn0013} yields
$ {\bf H}_1(1,2)=(\|{\bf a}(\theta_1)\|^2_2-{\rho_1}\|{\bf a}(\theta_0)\|^2_2)\cdot{\bf a}^{\mH}(\theta_0){\bf a}(\theta_1) $
and
$ {\bf H}_1(2,2)=\|{\bf a}(\theta_1)\|^4_2-{\rho_1}|{\bf a}^{\mH}(\theta_1){\bf a}(\theta_0)|^2 $.
Recalling \eqref{eqn0030}, we obtain
\begin{align}
{\bf c}_{\gamma}(1)=\dfrac{-{\Re}\left[{\bf a}^{\mH}(\theta_0){\bf a}(\theta_1)\right](\|{\bf a}(\theta_1)\|^2_2-{\rho_1}\|{\bf a}(\theta_0)\|^2_2)}{\|{\bf a}(\theta_1)\|^4_2-{\rho_1}|{\bf a}^{\mH}(\theta_1){\bf a}(\theta_0)|^2}.\nonumber
\end{align}
Meanwhile, since $ {\bf T}_0={\bf I} $, one obtains from \eqref{zetadef} that
$ \zeta={\rm sign}(\|{\bf a}(\theta_1)\|^2_2-{\rho_1}\|{\bf a}(\theta_0)\|^2_2) $.
Finally, from \eqref{eqn0361},
if $ {\rho_1}\leq\|{\bf a}(\theta_1)\|^2_2/\|{\bf a}(\theta_0)\|^2_2 $, we have
$ {\gamma}_{1,\star}={\gamma}_{1,a} $, otherwise,
we obtain $ {\gamma}_{1,\star}={\gamma}_{1,b} $.
Recalling $ {\mu}_{1,\star}={\gamma}_{1,a} $,
we complete the proof.

\section{Proof of \eqref{prooff1} and \eqref{prooff2}}
From the equivalence of \eqref{qua511} and \eqref{qua513}, one gets
$ {\bf b}_{k}=-({\bf I}+{\breve{\bf \Sigma}}_k{\bf A}^{\mH}_{k}{\bf A}_{k})^{-1}
{\breve{\bf \Sigma}}_k{\bf A}^{\mH}_{k}{\bf a}(\theta_0) $.
Multiplying by $ {\bf I}+{\breve{\bf \Sigma}}_k{\bf A}^{\mH}_{k}{\bf A}_{k} $ to both
sides from the left of this equality yields
$ {\breve{\bf \Sigma}}_k{\bf A}^{\mH}_k
{\left({\bf a}(\theta_0)+{\bf A}_{k}{\bf b}_{k}\right)}=-{\bf b}_{k} $.
Since $ {\breve{\bf \Sigma}}_k $ is a diagonal matrix
and $ {{\breve{\bf w}}_{k}}={\bf a}(\theta_0)+{\bf A}_{k}{\bf b}_{k} $, we obtain
\addtocounter{equation}{1}
\begin{align}\label{qua519}
{\breve{\bf \Sigma}}_k={\rm Diag}\left(-{\bf b}_{k}\oslash
({\bf A}^{\mH}_k{\breve{\bf w}}_{k})\right).
\end{align}

Furthermore, as $ {{\bf w}}_{k}\!=\!{{\bf w}}_{k-1}+{\mu}_k{\bf a}(\theta_k) $, $ {\bf b}_k=[{\bf b}^{\mT}_{k-1}~{\mu}_k]^{\mT} $ and
$ {\bf A}_k=[{\bf A}_{k-1}~~{\bf a}(\theta_k)] $,
we can rewrite \eqref{qua519} as
\eqref{eqn158} on the top of next page.
Consequently, the following formulation can be obtained:
\addtocounter{equation}{1}
\begin{align}\label{qua526}
\dfrac{1}{\breve{\beta}_{k,i}}\!=\!\left\{\!\!\!
\begin{array}{cc}
\dfrac{1}{\breve{\beta}_{k-1,i}}-
\dfrac{{\mu}_k{\bf a}^{\mH}(\theta_i){\bf a}(\theta_k)}{{\mu}_i},\!\!&1\leq i\leq k\!-\!1\\\\
-\dfrac{{\bf a}^{\mH}(\theta_k){\breve{\bf w}}_{k-1}}{{\mu}_k}-\|{\bf a}(\theta_k)\|^2_2,&i=k.
\end{array}
\right.
\end{align}
From \eqref{qua526}, the powers of interferences can be clearly observed.
After some calculation, either \eqref{prooff1} or \eqref{prooff2} can be derived.
This completes the proof.

\bibliography{FinalVersionPart1ofOPARC20171225}

\end{document}